\documentclass[prc,nofootinbib,showpacs,eqsecnum,byrevtex,floatfix,superscriptaddress,twocolumn]{revtex4}
\usepackage{graphics}
\usepackage{pstcol}
\usepackage{afterpage}
\usepackage[dutch,english]{babel}

\begin{document}

\title{Soft-core meson-baryon interactions. I. One-hadron-exchange potentials}
\author{H.\ Polinder}
\affiliation{Institute for Theoretical Physics, Radboud University Nijmegen,
         Nijmegen, The Netherlands }
\affiliation{Forschungszentrum J\"{u}lich, Institut f\"{u}r Kernphysik (Theorie), D-52425 J\"{u}lich, Germany}
\author{Th.A.\ Rijken}
\affiliation{Institute for Theoretical Physics, Radboud University Nijmegen,
         Nijmegen, The Netherlands } 
 
\date{version of: \today}

\begin{abstract}
The Nijmegen soft-core model for the pseudoscalar-meson baryon interaction is derived, analogous to the Nijmegen $NN$ and $YN$ models. The interaction Hamiltonians are defined and the resulting amplitudes for one-meson-exchange and one-baryon-exchange in momentum space are given for the general mass case. The partial wave projection is carried through and explicit expressions for the momentum space partial wave meson-baryon potentials are presented.
\end{abstract}
\pacs{12.39.Pn, 21.30.-x, 13.75.Gx, 13.75.Jz}

\maketitle

 \hyphenation{Po-lin-der}
 \hyphenation{Nij-me-gen}

\section{Introduction}
\label{sec:1}

The strong interactions between mesons and baryons, in particular for pion-nucleon, 
kaon-nucleon and antikaon-nucleon, have been the subject of investigation for some decades, 
experimentally as well as theoretically. 

A large number of scattering experiment have been performed to investigate the pion-nucleon 
interaction.
The empirical phase shifts are obtained from a partial wave (PW) analysis 
of the scattering observables, which analysis judges the consistency with general principles
of the scattering data, and which provides 
a compact representation of these data. 
Although at first sight in principle an infinite number of 
phase shifts need to be determined from the data, the strong interactions are short-ranged and 
only the lower partial wave phase shifts will suffice. In constructing theoretical models for 
these interactions, it is usually much more economic to use the results of a PW-analysis 
than the scattering observables themselves. 
Different pion-nucleon PW-analyses \cite{Arn95,Koch80,Car73}
give quite accurate and consistent results. 
The most recent pion-nucleon PW-analysis has been performed by Arndt et al. \cite{Arn95}, 
to which we refer for more information on the current pion-nucleon scattering data base.

However, the situation for the kaon-nucleon interaction is different from the 
pion-nucleon interaction. The kaon-nucleon scattering observables are known to less accuracy, 
especially at low energies, due to the relatively low flux of the kaon beams. 
Consequently, the different kaon-nucleon phase shift analyses do not give quite accurate 
and may be not totally consistent results. 
The most recent kaon-nucleon PW-analysis has been performed by Hyslop et al. \cite{Hys92}, 
where much information on the kaon-nucleon scattering data base can be found. 

This lack of empirical knowledge makes it impossible to construct realistic 
theoretical kaon-nucleon models, using as input only information from $KN$-data.

Recently there has been an increase of interest in the kaon-nucleon and antikaon-nucleon interaction. 
An exotic resonance, the so called 
``penta-quark'',
 in the isospin zero kaon-nucleon system has been observed 
\cite{Nak03}, this experiment, however, was not a simple scattering experiment and a resonance 
has never been seen in the present kaon-nucleon scattering data.

The construction of new {\it K-factories} at the Japan Proton Accelerator Research Complex 
(J-PARC), and at GSI (FAIR) in Germany, will hopefully change  
the experimental situation drastically.
One of the major beams of these new accelerators
will be kaon beams, having a much higher intensity (ca. ten times) than the presently 
available kaon beams, for example at Brookhaven National Laboratory and KEK. Therefore, 
in the near future many more and accurate experimental data on the kaon-nucleon and 
antikaon-nucleon interaction can therefore be expected.   
Other new scattering data could be delivered by the DA$\Phi$NE facility at 
Frascati \cite{Mai95}. These activities will give much stronger constraints 
on kaon-nucleon models and a better understanding of the role of $SU_f(3)$ in meson-baryon 
interactions. 
Akaishi and Yamazaki \cite{Aka02} have investigated the possibility of 
nuclear anti-kaon bound states in nuclei in the framework of the Brueckner-Hartree-Fock 
theory using a simple phenomenological antikaon-nucleon model.
Such a state has indeed been observed experimentally \cite{Iwa03}.

In view of these experimental and theoretical developments it is rather timely to construct 
theoretical kaon-nucleon models as realistically as possible, and this work is an attempt to do so. 

The subject of this work is the construction of a dynamical model for the pion-nucleon ($\pi N$) and kaon-nucleon ($K^+N$) interactions. 
In two papers we describe the so called Nijmegen soft-core meson-baryon model (NSC model) and report 
on the results, obtained so far.  
First a soft-core meson- and baryon-exchange model 
for the $\pi N$ interaction is derived, showing that the soft-core approach of the 
Nijmegen group is not only successful for baryon-baryon ($NN$ and $YN$) interactions but 
also for meson-baryon interactions. The rich and accurate $\pi N$ scattering data base 
is used to determine the nonstrange coupling constants. Several other $\pi N$ models 
already exist and the NSC $\pi N$-model, besides the value in its own right, 
mainly serves as a natural starting point for the construction of the NSC $K^+ N$-model. 
This $K^+N$-model is an $SU_f(3)$ extension of the NSC $\pi N$-model, similar to the 
successful Nijmegen soft-core one-boson-exchange nucleon-nucleon and hyperon-nucleon models 
\cite{NRS78} and \cite{MRS89}. In this way many parameters in the NSC $K^+ N$-model 
are determined by the NSC $\pi N$-model, and the lack of accurate $K^+N$-data can be 
overcome partially.

\begin{figure}[t]
\begin{center}
\resizebox{8.25cm}{10.86cm}{\includegraphics*[5cm,14.0cm][16cm,28.5cm]{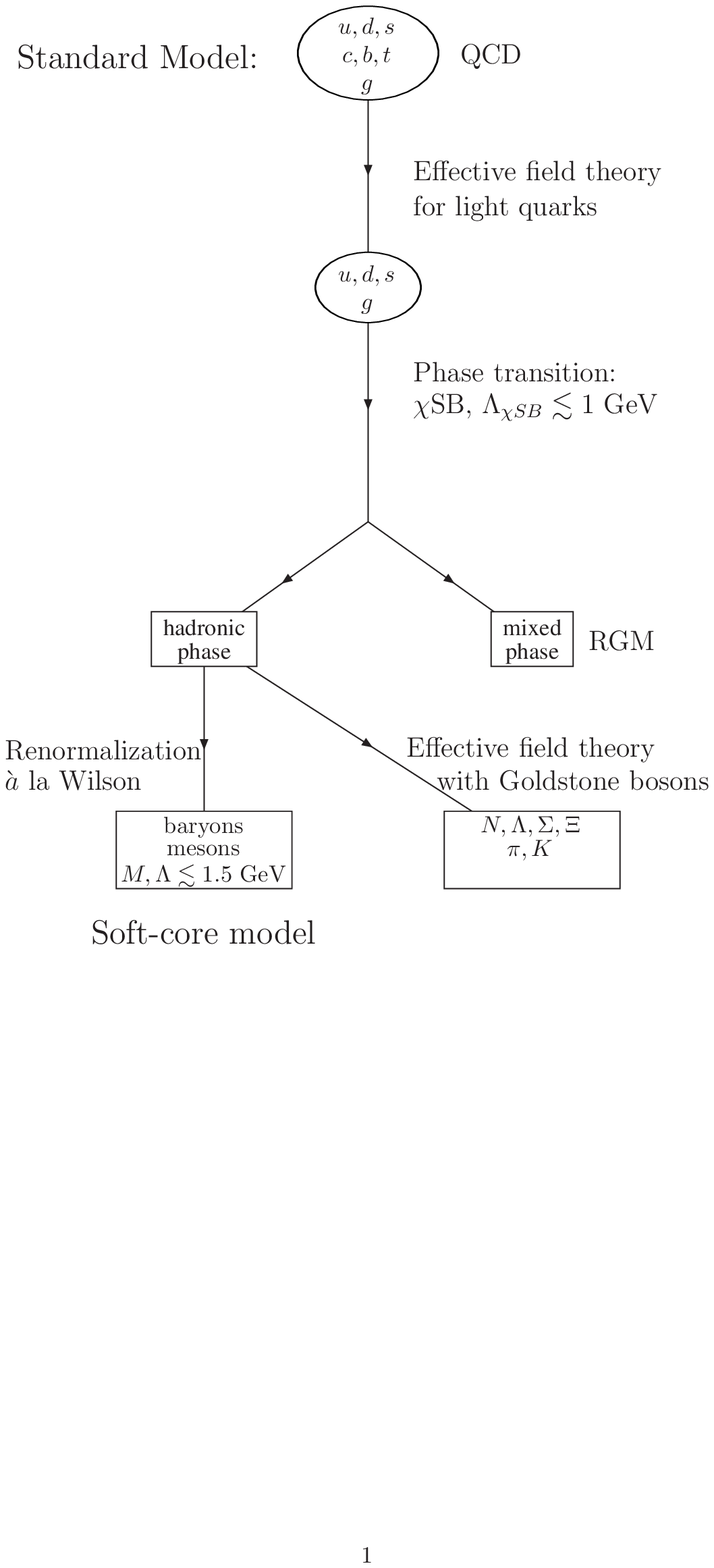}}
\end{center}
\caption{Overview of the theoretical basis for the soft-core meson-baryon interactions.}
\label{fig:0.0}
\end{figure}

In concept, the approach for the strong low- and intermediate-energy hadron-hadron interactions 
\cite{Wei79,Man84,Geo93}, used by the Nijmegen group, is schematically outlined in 
Figure \ref{fig:0.0}.
The starting point is the Standard Model, in which strong interactions occur between 
the six quarks and the gluons, and integrating out the heavier quarks to arrive at 
an effective QCD for the light quarks (u,d,s) only. 
Generally accepted, the vacuum of QCD becomes unstable for momenta transfer 
$q^2\le\Lambda^2_{\chi SB}\simeq 1 {\rm GeV}^2$ and the chiral symmetry is broken spontaneously 
($\chi$SB). The vacuum goes through a phase transition and generates constituent quark masses 
($m_q\approx 300$ MeV) and reduces the strong coupling constant $\alpha_s$. 
The pseudoscalar-mesons are viewed as the Nambu-Goldstone bosons originating from the 
$\chi {\rm SB}$, which makes it natural to assume the presence of a meson-cloud around 
the constituent quarks.

This provides a natural basis for an approach to the interaction between mesons and baryons 
using effective baryon-meson Lagrangians. 
At low- and intermediate-energies we do not consider a mixed phase of hadrons and quarks, as is 
done by others using the resonating-group method (RGM) \cite{Lib77,Rib80,Fuj96}, 
but restrict ourselves to the hadronic phase only. 
Furthermore, heavy baryons and mesons can be viewed as being integrated out, 
using for example the renormalization method in the manner of Wilson \cite{Pol84}, 
and an effective field theory, with meson and baryon masses with M$\le 1.5$ GeV, results.
In this work, this general picture is appealed to in the construction of a soft-core 
meson-baryon model for low- and intermediate-energy interactions.

In the NSC model the one-meson-exchange and one-baryon-exchange potentials 
are obtained from field theoretical Feynman diagrams in momentum space using effective interaction 
Hamiltonians, together with the meson-baryon Green's function they constitute the kernel of the 
two-particle integral equation for the amplitude, which is a three-dimensional reduction of the 
fully covariant (four-dimensional) Bethe-Salpeter equation \cite{Sal51}. 
Alternatively, one could view this work in the framework of the covariant perturbation theory 
as formulated by Kadyshevsky \cite{Kad64, Kad68a, Kad68b, Itz70}. Here the particles, also in the intermediate states,
remain on the mass shell, and pair-suppression can be implemented in a covariant way. Moreover, the 
three-dimensional integral equation obtained in the Kadyshevsky-scheme has exactly the 
same form as used in this work.

Form factors of the Gaussian type are introduced 
to take into account the extended size of the hadrons and to make the integral equation of 
the Fredholm type. The Coulomb interaction, playing a role at very low energies only and 
which is important in charge symmetry breaking (CSB) studies, will be neglected in this work.   
The integral equation for the amplitude is solved on the partial wave basis, in this way only 
one-dimensional integrals need to be performed to find the amplitude and the corresponding 
scattering observables or phase shifts for each partial wave.

We present this work in two papers. In general, this first paper, referred to as I,
contains a description of the theory and the second paper, referred to as paper II \cite{Pol05},  
gives the results for $\pi N$  and $K^+N$.

The contents of this first paper are as follows.
The definition of the field theoretical 
one-meson-exchange and one-baryon-exchange potentials in the context of a three-dimensional integral 
equation, a relativistic generalization of the Lippmann-Schwinger equation, is reviewed 
in Sec. \ref{chap:2}. We introduce the usual potential forms in Pauli-spinor space, 
where we include the central ($C$) and the spin-orbit ($SO$) potentials, 
which are the only relevant potentials in case of spin-1 spin-$\frac{1}{2}$ interactions. 
And the relations between the relativistic and center-of-mass amplitudes are given. 

The integral equation for the amplitude is solved on the partial wave basis in order 
to find the partial wave phase shifts, which are compared with the empirical phase shifts. 
Therefore we perform the basic partial wave projections, in particular those for the 
spinor invariants, in Sec. \ref{chap:3}. And the relations between the partial wave amplitudes, 
the phase shifts and the scattering observables, $\sigma$, $d\sigma/d\Omega$ and $P$, 
for spin-1 spin$\frac{1}{2}$ scattering are given. The partial wave basis is chosen according 
to the convention of \cite{SYM57}. 

The effective baryon-baryon-meson and meson-meson-meson interaction Hamiltonians from which 
the one-meson-exchange and one-baryon-exchange Feynman diagrams are derived, are given in 
Sec. \ref{chap:5}. The explicit expressions for the momentum space Feynman diagrams for 
scalar-meson-, vector-meson-, tensor-meson- and baryon-exchanges for general baryon and meson masses as well as their 
partial wave projections are also listed in this section.

 In the appendices added, details are given on the calculation of the partial wave matrix elements 
(Appendix \ref{app:AAA}), 
the one-particle-exchange Feynman diagrams (Appendix \ref{app:C}), and the expansion coefficients, 
$X$, $Y$ and $Z$, of the partial wave potentials in $x=\cos \theta$ for the different exchanges 
(Appendix \ref{app:I}).

For results and a discussion, we refer to paper II.
 
\section{The meson-baryon potential and amplitude}
\label{chap:2}
The field theoretical one-particle-exchange meson-baryon potentials in the context of a two-particle equation are defined in this section for the case of $J^{PC}=0^{-+}$ mesons and $J^P=\frac{1}{2}$ baryons. We approximate the Bethe-Salpeter equation by assuming ``dynamical pair suppression'', hence neglecting the propagation of negative-energy states and by integrating out the time variable, we end up with a three-dimensional integral equation for the meson-baryon amplitude in the center of mass system. The relations between the center of mass and the relativistic amplitudes are given in the last part of this section.
\subsection{Kinematics and relativistic amplitudes}
\label{sec:2}
We consider the meson-baryon or more specific the $\pi N$  and $K^+ N$ reactions
\begin{equation}
  {\mathcal P}_i(q)+{\mathcal B}_i(p,s) \rightarrow {\mathcal P}_f(q')+{\mathcal B}_f(p',s')\ ,
\label{eq:2.1} \end{equation}
where ${\mathcal P}$ stands for the pseudoscalar-mesons, and ${\mathcal B}$ stands for the baryons. We will refer to ${\mathcal P}_i$ and ${\mathcal P}_f$ as particles $1$ and $3$ and to ${\mathcal B}_i$ and ${\mathcal B}_f$ as particles $2$ and $4$. The four momentum of particle $i$ is $p_{i}=(E_{i},{\bf p}_{i})$ where $E_{i}=\sqrt{{\bf p}_{i}^{2}+M_{i}^{2}}$ is the energy and $M_{i}$ is the mass of particle $i$. In our convention the transition amplitude matrix $M$ is related to the $S$-matrix via
\begin{equation}
  \langle f|S| i\rangle= \langle f| i| \rangle -
  i(2\pi)^{4}\delta^{4}(P_{f}-P_{i})
  \langle f| M| i \rangle\ ,
\label{eq:2.2} \end{equation}
in this convention a negative potential corresponds with attraction and a positive potential with repulsion. Here $P_{i}=p+q$ and   $P_{f}=p'+q'$
represent the total four momentum for the initial state $|i\rangle$
and the final state $|f\rangle$. The latter refer to the
two-particle states, which we normalize in the following way, see e.g. \cite{Col68, Pil67},
\begin{eqnarray}
  \langle{\bf p'}_{1},{\bf p'}_{2}|{\bf p}_{1},{\bf p}_{2}\rangle
  &=&(2\pi)^{3}2E({\bf p}_{1})
  \delta^{3}({\bf p}_{1}'-{\bf p}_{1})
\times \nonumber \\ &&
(2\pi)^{3}2E({\bf p}_{2})
  \delta^{3}({\bf p}_{2}'-{\bf p}_{2})\ .
\label{eq:2.3}
\end{eqnarray}
With this normalization, the unpolarized differential cross section
in the center of mass (CM) system is given by
\begin{equation}
 \left(\frac{d\sigma}{d\Omega}\right)_{CM} = \frac{p_f}{p_i} \frac{1}{2} \sum
 \left|\frac{\langle f| M| i\rangle}{8\pi\sqrt{s}}\right|^2\ ,
\label{eq:2.4} \end{equation}
where $\sum$ stands for the summation over the spin of the final baryon.

Since in this work, the scattering particles are always on the
mass-shell, i.e. $p_i^2 = m_i^2$, parity conservation  and Lorentz invariance implies that 
the matrix elements of the $M$-operator for meson-baryon interactions, which is a $4\times 4$-matrix sandwiched between Dirac spinors, can be written in terms of two independent amplitudes
\begin{eqnarray}
  \langle f |M| i \rangle &=& 
\bar{u}_{B'}({\bf p'},s_f)\left[ \vphantom{\frac{A}{A}} 
A_{fi}(s,t,u) + \frac{ \not\! q' + \not\! q}{2} B_{fi}(s,t,u)\right]
\nonumber \\ &&\times \
 u_{B}({\bf p},s_i)\ ,
\label{eq:2.5} 
\end{eqnarray}
where $f$ and $i$ stand for the two-particle channels $\pi N, K^+ N$, etc....
In the Dirac spinors $s_f,s_i$ are the magnetic spin variables, which will be specified later.
The functions $A_{fi}(s,t,u)$ and $B_{fi}(s,t,u)$ are Lorentz scalars, and depend on the Mandelstam invariants
\begin{eqnarray}
s&=&\left(p+q\right)^2=\left(p'+q'\right)^2\ , \nonumber \\ 
t&=&\left(q'-q\right)^2=\left(p-p'\right)^2\ , \nonumber \\
u&=&\left(p-q'\right)^2=\left(p'-q\right)^2\ ,
\label{eq:2a.6} 
\end{eqnarray}
which satisfy the well-known (on-mass-shell) relation $s+t+u= \sum_{i=1}^4 m_i^2$.
The total and relative four-momenta ($P_c$ and $k_c$) of the initial,  final, and intermediate channel $(c=i,f,n)$ are defined by 
\begin{eqnarray}
  P_c &=& p_c + q_c\ ,\ k_c = \mu_{c,2}\ p_c - \mu_{c,1}\ q_c\ ,         
\label{eq:2b.1} \end{eqnarray}
where the weights are arbitrary besides the condition $\mu_{c,1}+\mu_{c,2}=1$. For each channel the four-momenta of the baryons and pseudoscalar-mesons ($p_c$ and $q_c$)  in terms of $P_c$ and $k_c$ are 
\begin{eqnarray}
  p_c &=& \mu_{c,1} P_c + k_c\ ,\ \ q_c = \mu_{c,2}\ P_c - k_c\ .  
\label{eq:2b.2} \end{eqnarray}
In this work we will use $\mu_{c,1}=\mu_{c,2}=\frac{1}{2}$. In the center-of-mass system (CM-system)
we have for on-mass-shell momenta
\begin{eqnarray}
p_c&=&\left(E\left({\bf p}_c\right),{\bf p}_c\right)\ ,\;q_c=\left({\cal E}\left({\bf p}_c\right),-{\bf p}_c\right)\ ,\nonumber \\
P_c&=&\left(W_c,{\bf 0}\right)\ ,\;\;\;\;\;\;\;\;\,k_c=\left(\mu_2E\left({\bf p}_c\right)-\mu_1{\cal E}\left({\bf p}_c\right),{\bf p}_c\right)\ ,
\nonumber \\
\end{eqnarray}
where the total energy is $W_c=\sqrt{s}=E({\bf p}_c)+{\cal E}({\bf p}_c)$.
Obviously the relative three-momentum is equal to the center-of-mass three-momentum of the baryon.

In general Feynman diagrams, in particular in the Green's functions, the particles are off-mass-shell. In the following the three-momenta for the initial and the final states are denoted respectively by ${\bf q}_i$ and ${\bf q}_f$. 
Because of translation invariance $P_i=P_f$ and so $\sqrt{s}=W_i=W_f$.
As introduced here,
the total energies in the CM-system are    
$W_i= E({\bf q}_i)+{\cal E}({\bf q}_i)$, and
$W_f= E({\bf q}_f)+{\cal E}({\bf q}_f)$.


\subsection{Relativistic two-particle equations}
\label{sec:2b}
The Bethe-Salpeter equation, a full two-particle relativistic scattering equation, for the ${\cal M}$-amplitudes reads
\begin{eqnarray}
\label{eq:2b.4} 
\lefteqn{
 {\cal M}_{fi}(q_f,q_i;P) = {\cal M}^{irr}_{fi}(q_f,q_i;P) +  \sum_{n}\int d^4k_n 
}
 \nonumber \\ && \times 
 {\cal M}^{irr}_{fn}(q_f,k_n;P)\: G(k_n,P)\: {\cal M}_{ni}(k_n,q_i;P)\, ,   
\end{eqnarray}
where the interaction kernel is denoted by ${\cal M}^{irr}$, and $G$ is the two-particle 
Green's function. The contributions to the kernel ${\cal M}^{irr}$ come from the meson-baryon
irreducible Feynman diagrams. The reducible diagrams are generated by the integral equation. In deriving Eq. (\ref{eq:2b.4}) the integration over the momenta
of the intermediate particles can be replaced by an integration over the total and relative momenta  $\int\int d^4p_n d^4q_n \rightarrow \int\int d^4P_n d^4k_n$. Then, using the conservation of the total four-momentum, one can perform $\int d^4P_c$ and separate an overall $(2\pi)^4\delta^4(P_f-P_i)$ factor.
\noindent The meson-baryon Green's function is given in terms of the one-particle Green's functions 
\begin{eqnarray}
 G(k_n,P) &=& \frac{i}{(2\pi)^4}
 \left[\frac{1}{\gamma\left(\mu_{n,1}P+k\right)-M_n}\right]^{(a)}
\nonumber \\ && \times
 \left[\frac{1}{\left(\mu_{n,2}P-k\right)^2-m_n^2}\right]^{(b)}.
\label{eq:2b.5} \end{eqnarray}
It is instructive to separate the positive- and the negative-energy components of the propagator.
For that purpose, we rewrite the one-particle propagators as follows. For the spin-$\frac{1}{2}$ baryons the off-mass-shell propagator can be written in terms of the Dirac spinors as  
\begin{equation}
\frac{\not\! p +M}{p^2-M^2+i\delta} = \frac{M}{ E({\bf p})}
\left[ \frac{\Lambda_+({\bf p})}{p^0- E({\bf p})+i\delta}-
\frac{\Lambda_-(-{\bf p})}{p^0+ E({\bf p})-i\delta} \right]\ ,
\label{eq:2b.6} \end{equation}
where the projection operators $\Lambda_+({\bf p})$ and $\Lambda_-({\bf p})$ on the
positive- and negative-energy states are \cite{Bjo65}
\begin{eqnarray}
\Lambda_+({\bf p})&=& \sum_s \frac{u({\bf p},s) \otimes \bar{u}({\bf p},s)}{2M}=\frac{\not\! p+M}{2M}\ , \nonumber \\
\Lambda_-({\bf p})&=&-\sum_s \frac{v({\bf p},s) \otimes \bar{v}({\bf p},s)}{2M}=\frac{-\not\! p+M}{2M}\ ,\ \ \ \ \
\label{eq:2b.7}
 \end{eqnarray}
and $u({\bf p},s)$ and $v({\bf p},s)$ are the Dirac spinors for spin-$\frac{1}{2}$ particles, which are on-mass-shell by definition. For the meson propagator similar to Eq. (\ref{eq:2b.6}) one has the identity        
\begin{equation}
\frac{1}{q^2-m^2+i\delta} = \frac{1}{2{\cal E}({\bf q})}\left[
\frac{1}{q^0-{\cal E}({\bf q})+i\delta}-\frac{1}{q^0+{\cal E}({\bf q})-i\delta} \right]\ .
\label{eq:2b.8} \end{equation}
Then, in the CM-system, where ${\bf P}=0$ and $P_{0}=W$, the meson-baryon Green's function 
can be written as
\begin{eqnarray}
 G(k_n,P) &=& \frac{i}{(2\pi)^4}\left[\frac{M}{2 E({\bf k}_n) {\cal E}({\bf k}_n)}\right] \nonumber \\
&&\times \left[ \frac{\Lambda_+({\bf k}_n)}{\mu_{n,1}W+k_n^0- E({\bf k}_n)+i\delta}
\right.\nonumber \\ &&\left.
-\frac{\Lambda_-(-{\bf k}_n)}{\mu_{n,1}W+k_n^0+ E({\bf k}_n)-i\delta} \right] \nonumber \\ 
&&\times \left[ \frac{1}{\mu_{n,2}W-k_n^0-{\cal E}({\bf k}_n)+i\delta}
\right.\nonumber \\ &&\left.
-\frac{1}{\mu_{n,2}W-k_n^0+{\cal E}({\bf k}_n)-i\delta} \right]\ . \label{eq:2b.9} 
\end{eqnarray}
Multiplying out Eq. (\ref{eq:2b.9}), writing the ensuing terms using an obvious
short hand notation, the contribution of the different propagating components
is displayed fully
\begin{eqnarray}
 G(k_n,P) &=& G^{+}(k_n,W) + G^{-}(k_n,W) 
\ , 
\label{eq:2b.10} \end{eqnarray}
where the superscripts $(+)$ and $(-)$ indicate the positive- and negative-energy baryon states.
 Considering similarly the amplitudes $M^{\beta,\alpha}_{ij}$:
\begin{eqnarray}
M_{ij}^{+,+}&=&\bar{u}_{B'}({\bf p}_f,s_f){\cal M}_{ij} u_{B}({\bf p}_i,s_i)\ , \nonumber \\
M_{ij}^{+,-}&=&\bar{u}_{B'}({\bf p}_f,s_f){\cal M}_{ij} v_{B}({\bf p}_i,s_i)\ , \nonumber \\
\rm{etc}. &&
\label{eq:2b.11} \end{eqnarray}
where the subscripts $i$ and $j$ refer to the different two-particle channels, one obtains from Eqs. (\ref{eq:2b.4}), (\ref{eq:2b.10}) and (\ref{eq:2b.11}) the full relativistic scattering equation
\begin{eqnarray}
 M_{fi}^{\beta,\alpha}(q_f,q_i;P) &=& (M^{irr})_{fi}^{\beta,\alpha}(q_f,q_i;P) +  
\nonumber \\ &&
 \sum_{n}\int d^4k_n
 (M^{irr})_{fn}^{\beta,\gamma}(q_f,k_n;P)\ \times 
 \nonumber \\ &&
 G_n^{\gamma}(k_n,P)\ M_{ni}^{\gamma,\alpha}(k_n,q_i;P)\ .
\label{eq:2b.12} \end{eqnarray}
In all we have $2^2=4$ amplitudes, which are coupled as illustrated in Eq. (\ref{eq:2b.12}).

The complexity of the previous equation can be reduced considerably if we assume dynamical pair-suppression, i.e. if we neglect the contribution of negative-energy states.
Then the full scattering equation Eq. (\ref{eq:2b.12}), for $\alpha=+$ and $\beta=+$, reduces to the four-dimensional integral equation
\begin{eqnarray}
 M_{fi}^{+,+}(q_f,q_i;P) &=& 
 (M^{irr})_{fi}^{+,+}(q_f,q_i;P) + 
 \nonumber \\ &&
\sum_n \int d^4k_n
 (M^{irr})_{fn}^{+,+}(q_f,k_n;P)\ 
 \nonumber \\ &&
G^{+}_n(k_n,P)\          
 M_{ni}^{+,+}(k_n,q_i;P)\ ,     
\label{eq:2b.14} \end{eqnarray}
with the positive-energy Green's function
\begin{eqnarray}
 G^{+}(k_n,P) &\approx& \frac{i}{(2\pi)^4}\left[
\frac{1}{4 E({\bf k}_n) {\cal E}({\bf k}_n)}\right] 
 \nonumber \\ && \times
\frac{1}{\left[\mu_1W+k_n^0- E({\bf k}_n)+i\delta\right]}
 \nonumber \\ && \times 
 \frac{1}{\left[\mu_2W-k_n^0-{\cal E}({\bf k}_n)+i\delta\right]}\ . 
\label{eq:2b.15} \end{eqnarray}
We note that this simplification in principle brings about a hopefully tolerable breach of relativistic invariance. On the other hand in Feynman diagrams particles go off-mass-shell, and the off-mass-shell behavior is not really known for mesons and baryons, certainly not if a truncated kernel is used, which is always the case. Then it might be better to allow positive-energy states only.
 
\subsection{Three-dimensional two-particle equations}
\label{sec:3}
Three-dimensional integral equations for the amplitudes can be derived in various ways. The methods assume 2-particle unitarity as a basic ingredient. The derivation for the meson-baryon systems follows the same procedure as that for the baryon-baryon channels. For the latter see e.g. references \cite{NRS77,Many1,Many2,Many3,Many4,GVS75,PVER}. In \cite{Rijken85} the derivation is based entirely on two-particle unitarity and the analyticity properties of the amplitudes, using the $N/D$-formalism. In the latter
approach the in essence Regge pole nature of meson-exchange can be
apprehended most easily. 

\subsubsection{On-mass-shell approximation}      
\label{sec:3.1}
\begin{figure}[t]
\begin{center}
\resizebox{8.25cm}{0.825cm}{\includegraphics*[1.2cm,23.5cm][16cm,25.cm]{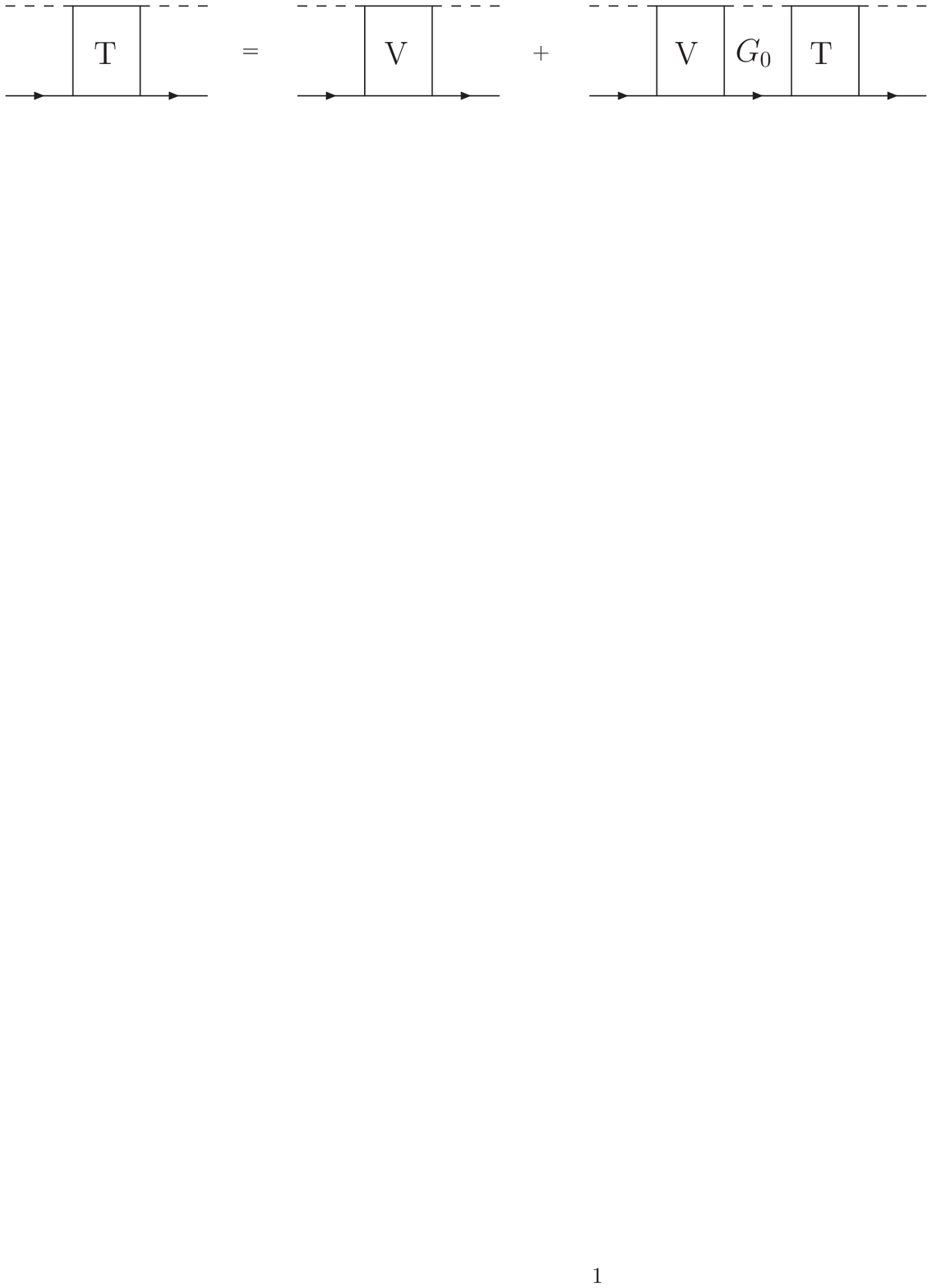}}
\end{center}
\caption{Diagrammatic representation of the meson-baryon scattering equation Eq. (\ref{eq:3.3}). The solid line denotes the baryon and the dashed line denotes the meson.}
\label{fig:2.0}
\end{figure}

The simplest way to reduce the four-dimensional integral equation, Eq. (\ref{eq:2b.14}),
to a three-dimensional one is to put the intermediate particles on the mass-shell, i.e. $p_n^0=E({\bf k}_n)=\sqrt{{\bf k}_n^2+M_n^2}$, $q_n^0={\cal E}({\bf k}_n)=\sqrt{{\bf k}_n^2+m_n^2}$. It can readily be shown from Eq. (\ref{eq:2b.2}) that the zero components of the relative and total momenta $k_n$ and $P_n$ are  given by
\begin{eqnarray}
k_n^0&=&\mu_{n,2}E({\bf k}_n)-\mu_{n,1}{\cal E}({\bf k}_n)\ ,\nonumber \\
P_n^0&=&E({\bf k}_n)+{\cal E}({\bf k}_n)\ .
\label{eq:3.1}
\end{eqnarray}
If we neglect the $k_n^0$-dependence of the amplitudes, and evaluate them at the value given by
Eq. (\ref{eq:3.1}), the dependence of the four-dimensional equation on $k_n^0$ only occurs in the Green's function, and the $k_n^0$-integration of the Green's function can be done. We can define the $k_n^0$-independent amplitudes
\begin{eqnarray}
T_{ni}({\bf k}_n,{\bf q}_i;W)&=&M_{ni}^{+,+}(\tilde{k}_n,q_i;P)\ , \nonumber \\
V_{fn}({\bf q}_f,{\bf k}_n;W)&=&\left(M^{irr}\right)_{fn}^{+,+}(q_f,\tilde{k}_n;P)\ ,
\end{eqnarray}
where $\tilde{k}_n^0=\mu_{n,2}E({\bf k}_n)-\mu_{n,1}{\cal E}({\bf k}_n)$. Now, the $k_n^0$-integration in Eq. (\ref{eq:2b.14}) can be carried through, this leads to
\begin{eqnarray}
G_0\left({\bf k}_n;W\right)&=&\int_{-\infty}^\infty dk_{n0}\ G_n^{+}(k_n;P) \nonumber \\
&=& \frac{1}{(2\pi)^3} \frac{1}{4E({\bf k}_n){\cal E}({\bf k}_n)}
\nonumber \\ && \times
\frac{1} {W-E({\bf k}_n)-{\cal E}({\bf k}_n)+i\delta}\ .
\label{eq:3.2} \end{eqnarray}
The four-dimensional integral equation, Eq. (\ref{eq:2b.14}), now results in the three-dimensional integral equation
, which is also derived in  \cite{PVER},
\begin{eqnarray}
\lefteqn{
T_{fi}({\bf q}_{f},{\bf q}_{i};W)=V_{fi}({\bf q}_{f},{\bf q}_{i};W) + \sum_{n}\int \frac{d^{3}k_{n}}{(2\pi)^{3}}\times
}
\nonumber \\&&
V_{fn}({\bf q}_{f},{\bf k}_{n};W)
G_{0}({\bf k}_{n},W)T_{ni}({\bf k}_{n},{\bf q}_{i};W)\  .
\label{eq:3.3}
\end{eqnarray}
The integral equation for the $T$-matrix, Eq. (\ref{eq:3.3}), is schematically given in Figure \ref{fig:2.0}.

We remark that the three-dimensional integral equation for the amplitude, Eq. (\ref{eq:3.3}), is here obtained as an approximation of the Bethe-Salpeter equation, but in the formulation of Quantum Field Theory (QFT) as developed by Kadyshevsky \cite{Kad64,Kad68a,Kad68b,Itz70} this integral equation is obtained without making any approximation. In this formulation of QFT all particles, in particular the intermediate particles, are always on the mass-shell in contrast to the formalism of Feynman. Hence a covariant form of pair-suppression can be introduced phenomenologically.

Until this subsection the intermediate particles were in principle off-mass-shell and the total four momentum was conserved. Now we have put the intermediate particles on-mass-shell, but now in principle they are off-energy-shell, which means that
 $W\neq E({\bf k}_n)+{\cal E}({\bf k}_n)$. And the total four momentum in not conserved, but the total three momentum is conserved.

{\bf Note} that if the intermediate state is on-energy-shell, i.e. $W=E({\bf k}_n)+{\cal E}({\bf k}_n)$, the two poles of the Green's function, Eq. (\ref{eq:2b.15}), coincide. The value of $k_n^0$ at which this 
``pinching'' occurs is given by the on-mass-shell value, Eq. (\ref{eq:3.1}),
\begin{equation}
k_n^0=\mu_{n,2}E({\bf k}_n)-\mu_{n,1}{\cal E}({\bf k}_n) \pm i\delta \ .
\end{equation}
The contribution to the integral around this value of $k_n^0$ will be dominant, due to the pinching of the poles. This is the rationale for the 
``on-mass-shell approximation''.\\

\subsubsection{Potentials for the 3-d integral equation}
In order to calculate cross sections or phase shifts we need to solve Eq. (\ref{eq:3.3}), which is a complex integral equation for the $T$-matrix, even for physical momenta. It is possible to transform Eq. (\ref{eq:3.3}) into a Lippmann-Schwinger equation (which can be Fourier-transformed into coordinate space). 
However we do our calculations always in momentum space, so we do not need to solve the Lippmann-Schwinger equation but we will always solve Eq. (\ref{eq:3.3}).
%

Using rotational invariance and parity conservation we expand the $T$-matrix, which is a $2\times 2$-matrix in Pauli-spinor space, into a complete set of Pauli-spinor invariants. Introducing the momentum vectors
\begin{equation}
  {\bf q}=\frac{1}{2}({\bf q}_{f}+{\bf q}_{i})\ ,\hspace{.5em}
  {\bf k}={\bf q}_{f}-{\bf q}_{i}\ ,\hspace{.5em}
  {\bf \hat{n}}={\bf \hat{q}}_{i}\times{\bf \hat{q}}_{f}
\ ,
\label{eq:3.20} \end{equation}
where ${\bf q}_{f}$ and ${\bf q}_{i}$ are the final and initial CM-three-momenta respectively, there are only two independent spinor invariants $P_{\alpha}$, 
rotational invariant and invariant under parity transformations.     
We choose for the operators $P_{\alpha}$ in spin-space
\begin{equation}
  P_{1}=1\ \ , \ \   P_{2}= \mbox{\boldmath $\sigma$} \cdot{\bf q}_i
  \times{\bf q}_f\ ,
\label{eq:3.21} \end{equation}
corresponding to the central and spin-orbit piece of the amplitude, now the expansion of the $T$-matrix in spinor invariants reads
\begin{eqnarray}
  T &=& \sum_{1}^{2} T_{\alpha}({\bf q}_f^2,{\bf q}_i^2,
  {\bf q}_f\cdot{\bf q}_i)\ P_{\alpha} \nonumber \\
 &=& f({\bf q}_f,{\bf q}_i) + i\ g({\bf q}_f,{\bf q}_i)\ 
 \left(\mbox{\boldmath $\sigma$}\cdot{\bf \hat{n}}\right)\ .
\label{eq:3.22} \end{eqnarray}
For the partial wave projection we found it convenient to rewrite 
the $T$-matrix in terms of the amplitudes $F$ and $G$
\begin{equation}
  T = F\left({\bf q}_f,{\bf q}_i\right) +
      (\mbox{\boldmath $\sigma$}\cdot\hat{\bf q}_f)\ 
      G\left({\bf q}_f,{\bf q}_i\right)\ 
  (\mbox{\boldmath $\sigma$}\cdot\hat{\bf q}_i)\ .
\label{eq:3.23} \end{equation}
The relation between the 
``spin-nonflip'' and 
``spin-flip'' amplitudes $f,g$ and the amplitudes $F,G$ is readily found to be
\begin{equation}
 F = f + (\hat{\bf q}_f\cdot\hat{\bf q}_i)\ g\ \ ,\ \ G= -g\ .
\label{eq:3.24} \end{equation}
The connection between the non-relativistic amplitudes $F$ and $G$ defined in Eq. (\ref{eq:3.23}) and the relativistic amplitudes $A$ and $B$ defined in Eq. (\ref{eq:2.5}) can be obtained in a straightforward way using the explicit representation of the Dirac spinors, as will be shown in Sec. \ref{sec:4}.
Similar to Eq. (\ref{eq:3.22}) we expand the potentials $V$, so
\begin{eqnarray}
  V &=& \sum_{1}^{2} V_{\alpha}({\bf q}_f^2,{\bf q}_i^2,
  {\bf q}_f\cdot{\bf q}_i)\ P_{\alpha} \nonumber \\
 &=& V_C({\bf q}_f,{\bf q}_i) + i\ V_{SO}({\bf q}_f,{\bf q}_i)\ 
\left(\mbox{\boldmath $\sigma$}\cdot {\bf n}\right)\ .
\label{eq:3.25} \end{eqnarray}

\subsubsection{Lippmann-Schwinger equation}
\label{sec:3.3}
In order to arrive at a Lippmann-Schwinger equation, one chooses a new Green's function $g({\bf k},W)$ which satisfies a dispersion relation in ${\bf p}^{2}(s)$ rather than in $s$ \cite{NRS77}. Then one obtains
\begin{equation}
  g({\bf k}_{n},W)=
  \frac{-1}{2[E({\bf k}_{n})+\mathcal{E}({\bf k}_{n})]}
  ({\bf k}_{n}^{2}-{\bf q}_{n}^{2}-i\delta)^{-1}\ ,
\label{eq:3.15} \end{equation}
where ${\bf q}_{n}$ is the on-energy-shell momentum. This Green's function is then used in the integral equation Eq. (\ref{eq:3.3}) instead of the Green's function \mbox{$G_{0}({\bf k}_{n},W)$}. So the corrections to $\langle f|W|i \rangle$ due to the  transformation of the Green's functions are neglected here, they are of higher order in the couplings and are usually discarded in an OBE-approach. With the substitution of $g$ for $G_0$, Eq. (\ref{eq:3.2}) becomes identical to Eq. (2.19) of \cite{NRS77}. From now on we follow Sec. II of \cite{NRS77} in detail, the transformation to the non-relativistic normalization of the two-particle states leads to states with
\begin{eqnarray}
 ({\bf p}_{1}',s_{1}';{\bf p}_{2}',s_{2}'|{\bf p}_{1},s_{1};
 {\bf p}_{2},s_{2} ) &=&
  (2\pi)^{6}\delta^{3}({\bf p}_{1}'-{\bf p}_{1}) \delta_{s_{1}',s_{1}}
\nonumber \\ &&\times
                 \delta^{3}({\bf p}_{2}'-{\bf p}_{2})
\delta_{s_{2}',s_{2}}\ .
\label{eq:3.16} \end{eqnarray}
For these states we define the non-relativistic ${\cal T}$-matrix
\begin{eqnarray}
  (f|{\cal T}|i) &=& \frac{1}{\sqrt{4\mu_{34}(E_{3}+E_{4})}}
  \langle f|T|i \rangle \frac{1}{\sqrt{4\mu_{12}(E_{1}+E_{2})}}\ ,
 \nonumber \\  
\label{eq:3.17} \end{eqnarray}
where $\mu_{12}$ and $\mu_{34}$ are the reduced masses for respectively
the initial and final state.
Then we get from Eq. (\ref{eq:3.3}) the Lippmann-Schwinger equation
\begin{eqnarray}
(3,4|{\cal T}|1,2) &=& (3,4|{\cal V}|1,2)\ + 
      \sum_{n}\int \frac{d^{3}k_{n}}{(2\pi)^{3}}
      (3,4|{\cal V}|n_{1},n_{2}) 
\nonumber \\ && \times
\frac{2\mu_{n_{1},n_{2}}}
 {{\bf q}_{n}^{2}-{\bf k}_{n}^{2}+i\delta} (n_{1},n_{2}|{\cal T}|1,2)\ ,
\label{eq:3.18} \end{eqnarray}
where the potential ${\cal V}$ is defined analogously to the ${\cal T}$-matrix, Eq. (\ref{eq:3.17}). 
If in the low-energy approximation, the energies are expanded in terms of the momenta squared, the Lippmann-Schwinger equation in momentum space can in principle be Fourier-transformed into the equivalent Schr\"{o}dinger equation in configuration space. However our calculations are always in momentum space, so we always solve Eq. (\ref{eq:3.3}).

\subsection{Relation between relativistic- and CM-amplitudes}
\label{sec:4}
The relation between the relativistic amplitudes $A$ and $B$ and the non-relativistic amplitudes $F$ and $G$ is found making use of the representation of the Dirac spinors \cite{Bjo65}.
Since in the three-dimensional integral equation, Eq. (\ref{eq:3.3}), off-energy shell amplitudes appear, we now distinguish between the CM-energies of the final and initial states, defined by
\begin{equation}
 W_f^2\equiv s_f = (p'+q')^2\ \ ,\ \ 
 W_i^2\equiv s_i = (p+q)^2\ .  
\label{eq:4.1} \end{equation}
Then, a straightforward calculation of the operators $1$ and $\not\!\! Q$ between Dirac spinors gives the corresponding operators between Pauli spinors.
\begin{eqnarray}
 \bar{u}({\bf p}_f,s_f)\, u({\bf p}_i,s_i) &=& \sqrt{(E_f+M_f)(E_i+M_i)}\ \chi^{\dagger}_f
\times \nonumber \\ &&
\left[ 1-\frac{
\mbox{\boldmath $\sigma$}\cdot{\bf p}_f\ \mbox{\boldmath $\sigma$}\cdot{\bf p}_i}
{(E_f+M_f)(E_i+M_i)} \right] \chi_i\ , \nonumber \\ 
&& \nonumber \\
 \bar{u}({\bf p}_f,s_f) \not\!\! Q\, u({\bf p}_i,s_i) &=& \sqrt{(E_f+M_f)(E_i+M_i)}\ \chi^{\dagger}_f 
\times \nonumber \\ &&
\left[\vphantom{\frac{\mbox{\boldmath $\sigma$}\cdot{\bf p}_f}{(E_f+M_f)}}
 \frac{1}{2}\left[ (W_f-M_f)+(W_i-M_i)\right] + \right. \nonumber \\
 && \frac{1}{2}\left[ (W_f+M_f)+(W_i+M_i)\right]\ 
\times  \nonumber \\ && \left.
\frac{\mbox{\boldmath $\sigma$}\cdot{\bf p}_f\ \mbox{\boldmath $\sigma$}\cdot{\bf p}_i}
{(E_f+M_f)(E_i+M_i)} \right]\chi_i\ ,
\label{eq:4.2} \end{eqnarray}
with
\begin{equation}
 Q^\mu = \frac{1}{2}\left( q_f + q_i\right)^\mu\ .  
\label{eq:4.3} \end{equation}
In Eq. (\ref{eq:4.2}) we used the shorthand notations $E_f =E({\bf p}_f)$ etc. for the
baryon variables. The meson variables were eliminated using $q^0 = W-E$ etc.
From the expressions in Eqs. (\ref{eq:4.2}), (\ref{eq:2.5}) and (\ref{eq:3.23})  we immediately obtain the relations between the amplitudes $F$, $G$ and $A$,$B$
\begin{eqnarray}
F({\bf p}_f,{\bf p}_i) &=& 
\sqrt{(E_f+M_f)(E_i+M_i)} \left[\vphantom{\frac{1}{2}} A(s,t,u)\ + \right. \nonumber \\
 && \left. \frac{W_f-M_f+W_i-M_i}{2}\, B(s,t,u)\right]\ , 
 \nonumber \\ && \nonumber \\
G({\bf p}_f,{\bf p}_i) &=& 
\sqrt{(E_f-M_f)(E_i-M_i)} \left[\vphantom{\frac{1}{2}}
 -A(s,t,u)\ + \right. \nonumber \\
 && \left. \frac{W_f+M_f+W_i+M_i}{2}\, B(s,t,u)\right]\ .           
\label{eq:4.4} \end{eqnarray}
 
\section{The partial wave equation}
\label{chap:3}
The NSC model is fitted to the partial wave analyses of the $\pi N$ and $K^+ N$ scattering data, for this purpose the integral equation for the meson-baryon amplitude must be solved on the partial wave basis. This section deals with the transformation of the integral equation on the plane wave basis to the integral equation on the partial wave (LSJ) basis. From the unitarity of the scattering matrix, the relation between the partial wave amplitude and the partial wave phase shifts is derived.
\subsection{Partial wave analysis}
\label{sec:5}
\footnote{In this section we use the non-relativistic normalization Eq. (\ref{eq:3.16}) of the two-particle states.}
The states for the meson-baryon system are characterized by $J,L$, where $J$ is the total angular momentum and $L$ the orbital angular momentum. The latter, for fixed $J$-value, can assume the values $L=J\mp \frac{1}{2}$, since the spin of the baryons is $S=\frac{1}{2}$. Distinguishing between the partial waves with parity $P=(-)^{J-1/2}$ and $P=(-)^{J+1/2}$, using rotational invariance, we can write the potential matrix elements on the LSJ-basis in the following way \\ \\
 (i) $P=(-)^{L_+},L_+=J-1/2$:
\begin{eqnarray}
\lefteqn{
  (q_f ; L' J' M'|\ V\ |q_i; L J M) =  
} \nonumber \\ &&
4\pi\ V^{J,L_+}(L',L)\ 
\delta_{J^{\prime}J} \delta_{M^{\prime}M} \delta_{L^{\prime}L}\ ,
\label{eq:5.1} \end{eqnarray}
\\
(ii) $P=(-)^{L_-},L_-=J+1/2$:
\begin{eqnarray}
\lefteqn{
  (q_f ; L' J' M'|\ V\ |q_i; L J M) =  
} \nonumber \\ &&
4\pi\ V^{J,L_-}(L',L)\ 
\delta_{J^{\prime}J} \delta_{M^{\prime}M} \delta_{L^{\prime}L}\ .
\label{eq:5.2} \end{eqnarray}
Because of parity conservation in strong interactions, the $L_+=J-1/2$ and the $L_-=J+1/2$ waves obviously are decoupled. So mixing between states with different angular momentum never occurs.

The spherical wave functions in momentum space with quantum numbers $J,M,L,S=1/2$ are
\begin{equation}
{\cal Y}^M_{JL}(\hat{p},s) = 
 \sum_{m,\mu}\ C^{L\ \frac{1}{2}\ \, J}_{m\  \mu\  M}\ 
 Y^L_m(\hat{p})\ \chi^{\left(\frac{1}{2}\right)}_\mu(s)\ ,
\label{eq:5.3} \end{equation}
where $s$ is a spin variable for the baryons. For example, $s$ denotes the helicity of the baryon, or the projection of the spin along the normal $\hat{\bf n}$ to the scattering plane, or the projection of the spin along the $z$-axis. The latter spin variable we will use in this work. Then, in Eq. (\ref{eq:5.3}) we have $\chi^{(1/2)}_\mu(s) = \delta_{s,\mu}$. The central and non-central potential matrix elements on the LSJ-basis are derived in detail in Appendix \ref{app:bB}, the results are\\

\noindent 1.\ \textit{Central} $P_1=1$:
\begin{eqnarray}
 \lefteqn{\left(q_f;L'J'M'|F({\bf q}_f,{\bf q}_i)|q_i;LJM\right) =}
\nonumber \\
&&4\pi F_L\left(q_f,q_i\right)\  \delta_{L'L} \delta_{J',J} \delta_{M',M}\ ,
\label{eq:5.4} \end{eqnarray}

\noindent 2.\ \textit{ Non-central} 
  $P'_{2}= (\mbox{\boldmath $\sigma$}\cdot\hat{\bf q}_f)
  (\mbox{\boldmath $\sigma$}\cdot\hat{\bf q}_i)$:   
\begin{eqnarray}
\lefteqn{\left(q_f;L'J'M'|G({\bf q}_f,{\bf q}_i)|q_i;LJM\right) =}
\nonumber \\
&&4\pi \sum_{L''}\ 
 a_{L',L''} G_{L''}\left(q_f,q_i\right) a_{L'',L}\ 
\delta_{J',J}\delta_{M',M}\ ,
\label{eq:5.9} \end{eqnarray}
where the partial wave projections $F_L$ and $G_L$ as well as the matrix $a_{L',L}$ are defined in Appendix \ref{app:AAA}. The partial wave potentials $V^{J,L_+}$ and $V^{J,L_-}$ in Eqs. (\ref{eq:5.1}) and (\ref{eq:5.2}) can be expressed in terms of the partial wave expansions of $F$ and $G$. 
As expected from parity conservation, the partial wave potentials are diagonal in the $L=(J\pm 1/2)$-space
\begin{eqnarray}
V^{J,L_{\pm}}&=&F_{L_{\pm}}+G_{L_{\pm}\pm1}\ .
\label{eq:5.17a}
\end{eqnarray}
The partial wave potentials can also be expressed in terms of the partial wave projections of the central and spin-orbit potential. The relation between $F$ and $G$ and the central and spin-orbit potentials is given by Eq. (\ref{eq:3.24}),
\begin{eqnarray}
F&=&V^C+\cos \theta\ V^{SO}\ , \ G=-V^{SO}\ .
\end{eqnarray}
The partial wave potentials in terms of the partial wave projections of the central and spin-orbit potentials becomes
\begin{eqnarray}
\label{eq:5.17b}
V^{J,L_+}&=&V_{L_+}^C\ ;  \hspace{3.2cm} L_+=0\ , \nonumber \\
V^{J,L_{\pm}}&=&V_{L_{\pm}}^C+\frac{L_{\pm}+1}{2L_{\pm}+1}V_{L_{\pm}+1}^{SO}+
\nonumber \\ &&
\frac{L_{\pm}}{2L_{\pm}+1}V_{L_{\pm}-1}^{SO}-V_{L_{\pm}\pm1}^{SO}\ ; \hspace{.2cm}  L_{\pm}\geq1\  . 
\end{eqnarray}

\subsection{Partial wave integral equations and the unitarity relations}
\label{sec:6}
\subsubsection{Partial wave integral equations}
\label{sec:6.1}
The integral equation Eq. (\ref{eq:3.3}) we write first explicitly in terms of the plane wave states
\begin{eqnarray}
\lefteqn{
\left({\bf q}_{f},s_f|T(\sqrt{s})|{\bf q}_{i},s_i\right) =
 \left({\bf q}_{f},s_f|V(\sqrt{s})|{\bf q}_{i},s\right)\ }
\nonumber \\ &&
+ \sum_{n}\int\!\! \frac{d^{3}k_{n}}{(2\pi)^{3}} 
  \left({\bf q}_{f},s_f|V(\sqrt{s})|{\bf k}_{n},s_n\right) \ G_{0}({\bf k}_{n},\sqrt{s})
\nonumber \\ && \times
  \left({\bf k}_{n},s_n|T(\sqrt{s})|{\bf q}_{i},s_i\right)\ ,
\label{eq:6.1a} \end{eqnarray}
where, apart from spin-space, the amplitude $T$, the Green's function $G_0$ and the potential $V$ are matrices in the two-particle channel space. The partial wave $T$-matrix for $L=L_i=L_f$ defined by
\begin{equation}
 T_{J,L}(q_f,q_i;\sqrt{s}) = \left( q_f;LJM|T(\sqrt{s})|q_i;LJM\right)\  ,
\label{eq:6.2a} \end{equation}
which is independent of $J_z=M$ due to rotation invariance, is related to the $T$-matrix on the plane wave basis by
\begin{eqnarray}
\lefteqn{
T_{J,L}(q_f,q_i;\sqrt{s}) = \sum_{s_f,s_i} \!\int\!\frac{d^3 q'_f}{(2\pi)^3}\! \int\!\frac{d^3 q'_i}{(2\pi)^3} \left( q_f;LJM|{\bf q}'_f,s_f\right) }
\nonumber \\ && \times
\left( {\bf q'}_f,s_f|T(\sqrt{s})|{\bf q'}_i,s_i\right) \left( {\bf q}'_i,s_i| q_i;LJM\right)\ . \hspace{1.cm}
\label{eq:6.3} \end{eqnarray}
The integral equation for the partial wave amplitude now becomes
\begin{eqnarray}\lefteqn{
 T_{J,L}(q_f,q_i;\sqrt{s}) = V_{J,L}(q_f,q_i;\sqrt{s}) + 
\sum_{n}\int_{0}^{\infty}\ \frac{k_n^2 dk_n}{(2\pi)^3}\times
} \nonumber \\&&
 V_{J,L}(q_f,q_n;\sqrt{s})\ G_{0}(k_{n},\sqrt{s})
 T_{J,L}(q_n,q_i;\sqrt{s})\ .
\hspace{.4cm}                                       
\label{eq:6.7} \end{eqnarray}

\subsubsection{Partial wave unitarity relations, phase shifts}
\label{sec:6.2}
From the unitarity of the $S$-matrix, $S^\dagger\ S=1$, the $M$-matrix in Eq. (\ref{eq:2.2}) satisfies the condition
\begin{eqnarray}
2 \Im \langle f|M|i\rangle &=&
-(2\pi)^4\ \sum_n\ \delta^4(P_f-P_n)\
\nonumber \\ && \times
 \langle f| M^\dagger|n\rangle \langle n|M|i\rangle\ .
\label{eq:6.8} \end{eqnarray}
In deriving Eq. (\ref{eq:6.8}) one factors out a $\delta^4(P_f-P_i)$.
The previous equation for the CM-amplitudes can be written more explicitly, see for example Eq. (II.1.14) of \cite{PVER}, as
\begin{eqnarray}
\label{eq:6.9} 
\lefteqn{
 2 \Im \left({\bf q}_f,s_f|T|{\bf q}_i,s_i\right)
 =  -\frac{1}{(2\pi)^2}\sum_{n}\int\frac{d^3k_n}{4E({\bf k}_n){\cal E} ({\bf k}_n)} 
}\nonumber \\ && \times
  \left({\bf q}_f,s_f|T^\dagger|{\bf k}_n,s_n\right)
\delta(\sqrt{s}-E({\bf k}_n)-{\cal E}({\bf k}_n)) 
\nonumber \\ && \times
  \left({\bf k}_n,s_n|T^{\vphantom{A}}|{\bf q}_i,s_i\right)\ ,
\end{eqnarray}
where the summation $\sum_n$ is over all intermediate two-particle channels coupled to the initial and final state. Here, $s_i$, $s_f$, and $s_n$ are the spin labels for the initial, final, and intermediate states.
The momentum of the intermediate state $k_n$ is such that $E({\bf k}_n)+{\cal E}({\bf k}_n) = \sqrt{s}$. 
The unitarity relation for the partial wave amplitude then becomes
\begin{eqnarray}
 2 \Im T_{J,L}(q_f,q_i) &=& -\sum_{n}\frac{q_n}{16\pi^2\sqrt{s}} T^\dagger_{J,L}\left(q_f,q_n\right)
\nonumber \\ && \times
T_{J,L}\left(q_n,q_i\right)\ ,
\label{eq:6.10} \end{eqnarray}
where $q_n$ is the on-energy-shell momentum of the intermediate state. Introducing the partial wave amplitudes $F_{J,L}$ by the definition
\begin{equation}
 T_{J,L} = -32\pi^2\sqrt{s}\ F_{J,L}\ ,
\label{eq:6.11} \end{equation}
we find the simple unitarity relation for these amplitudes
\begin{equation}
\Im F_{J,L}(q_f,q_i) = \sum_c q_c\ F_{J,L}(q_f,q_c)^\dagger F_{J,L}(q_c,q_i)\ .
\end{equation}
For the single channel case $q_c=q_i=q_f=q$, phase shifts can be defined for the partial wave amplitude $F_{J,L}$ in the usual way
\begin{equation}
 F_{J,L} = \frac{1}{q}\ \sin\delta_{J,L}(q) \exp\left( i\delta_{J,L}(q)\right)\ .
\label{eq:6.12} \end{equation}
The relation of $F_{J,L}$ with the partial wave $S$-matrix is
\begin{equation}
 S_{J,L}=e^{2i\delta_{J,L}} = 1 + 2iq\ F_{J,L}\ .
\label{eq:6.13} \end{equation}
Now the expression for the differential cross section becomes
\begin{eqnarray}
\frac{d\sigma}{d\Omega}&=&\left|\tilde{f}\right|^2+\left|\tilde{g}\right|^2\ ,
\label{eq:6.14a}
\end{eqnarray}
where the commonly used spin-nonflip and spin-flip amplitudes $\tilde{f}$ and $\tilde{g}$ are given by
\begin{eqnarray}
\tilde{f}&=&\frac{f}{8\pi\sqrt{s}}=\sum_{L}\left[\left(L+1\right)F_{L+\frac{1}{2},L}+LF_{L-\frac{1}{2},L}\right]P_L(\cos\theta)\ , \nonumber \\
\tilde{g}&=&\frac{g}{8\pi\sqrt{s}}=\sum_{L}\left[F_{L+\frac{1}{2},L}-F_{L-\frac{1}{2},L}\right]\sin\theta \ \frac{dP_L(\cos\theta)}{d\cos\theta}\ .
\nonumber\\&&
\label{eq:6.14b}
\end{eqnarray}
The expressions for the total cross section, which is found by integrating the differential cross section, and the polarization are
\begin{eqnarray}
\sigma&=&4\pi\sum_J\frac{2J+1}{2}\left(\left|F_{J,L_+}\right|^2+\left|F_{J,L_-}\right|^2\right)\ , \nonumber \\
P(\theta)&=&\frac{2\Im \left(\tilde{f}\tilde{g}^*\right)}{\left|\tilde{f}\right|^2+\left|\tilde{g}\right|^2}\ .  
\label{eq:6.14c}
\end{eqnarray}


\section{Baryon- and meson-exchange potentials}
\label{chap:5}
The effective local interaction Hamiltonians that are used to calculate the one-hadron-exchange potentials are defined in this section. The Lorentz structure of the interaction is given and the $SU_f(3)$ structure is reviewed, since we extend the NSC $\pi N$-model to the NSC $K^+ N$-model. The amplitudes of the one-hadron-exchange Feynman diagrams are given and a partial wave projection is made to find the partial wave potentials.

\subsection{The interaction Hamiltonians}
\label{sec:7.98}
The potentials we use are obtained from the $t$-channel one-meson-exchange (OBE) and the $u$- and $s$-channel baryon-exchange Feynman diagrams. In the $t$-channel we consider the exchange of vector- and scalar-mesons and in the $u$- and $s$-channel we consider the exchange of $J^P=\frac{1}{2}^+$ and $\frac{3}{2}^+$-baryons.  In this work we also include Pomeron-exchange diagrams, where the physical nature of the Pomeron can be understood in the light of QCD as a two-gluon-exchange effect, see \cite{LOW75, NUS75}. The contribution of the Pomeron will almost completely cancel the contribution of the isoscalar scalar-meson $\sigma$.

The OBE Feynman diagrams for meson-baryon interactions contain a meson-baryon-baryon vertex and a meson-meson-meson vertex. These vertices are determined by the effective local interaction Hamiltonian densities. The Lorentz structure of the local interaction densities for the meson-baryon-baryon (MBB) vertices we use are listed below
\begin{itemize}
\item[a.] \underline{$J^{PC}=0^{-+}$ Pseudoscalar-mesons}:\\
For the pseudoscalar-mesons we use the pseudovector interaction Hamiltonian
\begin{equation}
 {\cal H}_{PV} = \frac{f}{m_{\pi^+}}\ \bar{\psi}_f\gamma_5\gamma_\mu\psi_i\
\partial ^{\mu} \phi_P\ ,
\label{eq:7.1} \end{equation}
which is scaled with the charged-pion mass in order to have a dimensionless pseudovector coupling constant.
\item[b.] \underline{$J^{PC}=1^{--}$ Vector-mesons}:\\
The interaction Hamiltonian is given in terms of the electric and magnetic interaction
\begin{eqnarray}
 {\cal H}_{V} &=& g_V\ \bar{\psi}_f\gamma_\mu\psi_i\ \phi^\mu_V
\nonumber \\ &&
+ \frac{f_V}{4{\cal M}}\bar{\psi}_f \sigma_{\mu\nu}\psi_i 
 \left(\partial^\mu\phi^\nu_V-\partial^\nu\phi^\mu_V\right)\ ,
\label{eq:7.2} \end{eqnarray}
where usually the proton mass is used for ${\cal M}$ to scale the magnetic part of the interaction Hamiltonian. The antisymmetric tensor operator used here, is defined as $\sigma_{\mu\nu}=\frac{i}{2}\left[\gamma_{\mu},\gamma_{\nu}\right]$.
\item[c.] \underline{$J^{PC}=0^{++}$ Scalar-mesons}:
\begin{equation}
 {\cal H}_{S} = g_S\ \bar{\psi}_f\psi_i\ \phi_S\ .
\label{eq:7.3b} \end{equation}
Since we include Pomeron-exchange in the NSC model, the scalar-meson-exchange is canceled for the greater part, hence it is possible to satisfy the soft-pion theorem while including scalar-meson-exchange.
\item[d.] \underline{$J^{PC}=2^{++}$ Tensor-mesons}: \\
\noindent For the tensor-mesons we use the interaction Hamiltonian
\begin{eqnarray}
 {\cal H}_{T} &=& \left[\frac{i}{4}\bar{\psi}_f\left(
 \gamma_\mu \stackrel{\leftrightarrow}{\partial_\nu} +
 \gamma_\nu \stackrel{\leftrightarrow}{\partial_\mu} 
 \right)\psi_i\ F_1 -
\right. \nonumber \\ && \left.
\frac{1}{4}\left(\bar{\psi}_f\
 \stackrel{\leftrightarrow}{\partial^\mu} 
 \stackrel{\leftrightarrow}{\partial^\nu} 
 \psi_i\right)\ F_2 \right]\ \phi^{\mu\nu}_T\ ,
\label{eq:7.4b} \end{eqnarray}
where the coupling constants $F_1$ and $F_2$ are related to the dimensionless Pauli coupling constants by $G_{T,1}={\mathcal M}F_1$ and $G_{T,2}={\mathcal M}^2F_2$. Using the Gordon decomposition the Pauli coupling constants are related to the Dirac coupling constants by $g_T=G_{T,1}+G_{T,2}$ and $f_T=-G_{T,2}$.
\item[e.] \underline{$J^P=\frac{3}{2}^+$ Resonance-Baryon-Pseudoscalar-meson:} \\
\noindent The local interaction density for the $J^P=\frac{3}{2}^+$ Resonance -Nucleon-Pseudoscalar-meson  ($Y^* N P$) interaction is
\begin{equation}
{\cal H}_{Y^* NP} = -i \frac{f^*_{}}{m_{\pi^+}}\ \bar{\psi}_N\ 
 \psi_{Y^*,\mu}\ \partial^\mu\phi_P\ ,
\label{eq:7.5} \end{equation}
where the charged-pion mass makes the coupling dimensionless. We use the Rarita-Schwinger formalism for the spin-3/2 resonances, see
e.g. \cite{Pil67,Car71}.
\item[f.] \underline{$J^P=\frac{1}{2}^-$ Resonance-Baryon-Pseudoscalar-meson:} \\
The local interaction Hamiltonian for the $J^P=\frac{1}{2}^-$ Resonance-Nucleon-Pseudoscalar-meson ($RNP$) interaction is 
\begin{equation}
 {\cal H}_{RNP} = \frac{f^{*(v)}_{}}{m_{\pi^+}}\ \bar{\psi}_N\gamma_\mu\psi_R\
\partial ^{\mu} \phi_P\ ,
\label{eq:7.1a} \end{equation}
where $\psi_R$ denotes the $J^P=\frac{1}{2}^-$ resonance, which has opposite parity to the nucleon. The $J^P=\frac{1}{2}^-$ resonances we consider in this work  are the $S_{11}(1555)$ in the $\pi N$ system and the $\Lambda (1405)$ in the $KN$ system.
\end{itemize}
Here $\phi$ denotes the pseudoscalar-, vector-, scalar-, and tensor-meson fields respectively  and $\psi$ denotes the baryon fields. The Pomeron-baryon-baryon interaction density we use, has the same Lorentz structure as the scalar-mesons.

We note that, making use of the Dirac equation $\left(\gamma^{\mu}\partial_{\mu}+M\right)\psi=0$, the pseudovector interaction Hamiltonian density in Eq. (\ref{eq:7.1}) is
``equivalent'' to the pseudoscalar density ${\cal H}_{PS} = i g\ \bar{\psi}_f\gamma_5\psi_i\ \phi_P$ for on-mass-shell particles. The coupling constants are then related according to
$ g/(M_{B_f}+M_{B_i}) = f/m_{\pi^+}$.
Analogous we find that the vector coupling Hamiltonian density in Eq. (\ref{eq:7.1a}) is 
``equivalent'' to the scalar density ${\cal H}_{S} = ig^{*(s)} \bar{\psi}_N\psi_R\ \phi_P$ for on-mass-shell particles. The coupling constants are in this case  related according to
$ g^{*(s)}/(M_N-M_R) = f^{*(v)}/m_{\pi^+}$.

The Lorentz structure of the local interaction density for triple-meson (MMM) vertices is schematically given below, they are discussed in more detail in paper II. 
\begin{itemize}
\item[(i)] \underline{$J^{PC}=1^{--}$ Vector-mesons}:
\begin{eqnarray}
{\cal H}_{PPV} &=& g_{PPV}\  \phi_V^\mu\ \left(\phi_P \stackrel{\leftrightarrow}{\partial}_\mu
 \phi_P\right) \ .
\label{eq:7.6} 
\end{eqnarray}
\item[(ii)] \underline{$J^{PC}=0^{++}$ Scalar-mesons}:
\begin{eqnarray}
{\cal H}_{PPS}& = &g_{PPS}\ \phi_S\ \left(\phi_P\ \phi_P\right) \ .
\label{eq:7.7} 
\end{eqnarray}
\item[(iii)] \underline{$J^{PC}=2^{++}$ Tensor-mesons}:
\begin{eqnarray}
{\cal H}_{PPT} &= &\frac{2g_{PPT}}{m_{\pi^+}}\ \phi_T^{\mu\nu} 
\left(\partial_\mu \phi_P\right)\left(\partial_\nu \phi_P\right)\ .
\label{eq:7.8} 
\end{eqnarray}
\end{itemize}
Concerning the flavor structure of the interaction densities, we assume that the coupling constants are related via $SU_f(3)$ symmetry, as outlined in paper II
, here the relevant isoscalar and isospin factors are given. However the potentials will break the $SU_f(3)$ symmetry dynamically, since we use the physical masses of the particles. 

\subsection{The relativistic invariant amplitudes}
\label{sec:7.99}
Using the previously defined interaction Hamiltonians, we give, besides the isospin and isoscalar factors, the contributions to the relativistic invariant amplitudes $A(s,t,u)$ and $B(s,t,u)$ in Eq. (\ref{eq:2.5}) for the elastic (e.g. $\pi N$ and $K^+ N$) channels, i.e.  $M_i = M_f \equiv M$, $ m_i = m_f = m$, where $M_f$ and $M_I$ are the final and initial baryon masses and $m_f$ and $m_i$ are the final and initial pseudoscalar-meson masses respectively. 
Amplitudes for the general mass case are listed in Appendix \ref{app:C}. 

\subsubsection{Baryon-exchange amplitudes}
 For $J^P=\frac{1}{2}^+$ baryon-exchange the relativistic amplitudes are
\begin{eqnarray}
 A_{ps}(s,t,u) &=& -\frac{g_{14} g_{23}}{ u- M_B^2 + i\epsilon}\left[ M_B - M\right] , \nonumber \\
 B_{ps}(s,t,u) &=& -\frac{g_{14} g_{23}}{u- M_B^2 + i\epsilon}\ ,
\nonumber \\
 A_{pv}(s,t,u) &=& -\frac{f_{14} f_{23}/m_{\pi^+}^2}{u-M_B^2+i\epsilon}\left[\vphantom{M^2M_B} u\left(M+M_B\right)
\right.\nonumber \\
&&\left.
-M^3 -M^2M_B\right]\ , \nonumber \\
B_{pv}(s,t,u) &=& -\frac{f_{14} f_{23}/m_{\pi^+}^2}{u-M_B^2+i\epsilon}\left[u+2MM_B
\right. \nonumber \\ && \left.
+M^2\right]\ , 
\label{eq:7.10} 
\end{eqnarray}
for pseudoscalar ($ps$) and pseudovector ($pv$) coupling respectively, $M_B$ is the mass of the exchanged baryon. The $J^P=\frac{1}{2}^+$ baryon direct pole gives rise to the relativistic amplitudes

\begin{eqnarray}
 A_{ps}(s,t,u) &=& -\frac{g_{12} g_{34}}{s- M_B^2 + i\epsilon}\left[M_B - M \right], \nonumber \\
 B_{ps}(s,t,u) &=& \frac{g_{12} g_{34}}{s- M_B^2 + i\epsilon}\ ,
\nonumber \\
 A_{pv}(s,t,u) &=& -\frac{f_{12} f_{34}/m_{\pi^+}^2}{s-M_B^2+i\epsilon}\left[\vphantom{M^2M_B} s\left(M+M_B\right)
\right.\nonumber \\
&&\left.
-M^3 -M^2M_B\right]\ , \nonumber \\
B_{pv}(s,t,u) &=& \frac{f_{12} f_{34}/m_{\pi^+}^2}{s-M_B^2+i\epsilon}\left[s+2MM_B
\right. \nonumber \\ && \left.
+M^2\right]\ , 
\label{eq:7.16}
\end{eqnarray}
for pseudoscalar and pseudovector coupling respectively.
For $J^P=\frac{1}{2}^-$ baryon-exchange the relativistic amplitudes are
\begin{eqnarray}
 A_{s}(s,t,u) &=& \frac{g^{*(s)}_{14} g^{*(s)}_{23}}{u- M_B^2 + i\epsilon}\left[M_B + M \right] , \nonumber \\
 B_{s}(s,t,u) &=& -\frac{g^{*(s)}_{14} g^{*(s)}_{23}}{u- M_B^2 + i\epsilon}\ ,
\nonumber \\
 A_{v}(s,t,u) &=& \frac{f^{*(v)}_{14} f^{*(v)}_{23}/m_{\pi^+}^2}{u-M_B^2+i\epsilon}\left[\vphantom{M^2M_B} u\left(-M+M_B\right)
\right. \nonumber \\ && \left.
+M^3 -M^2M_B\right]\ , \nonumber \\
B_{v}(s,t,u) &=& \frac{f^{*(v)}_{14} f^{*(v)}_{23}/m_{\pi^+}^2}{u-M_B^2+i\epsilon}\left[-u+2MM_B
\right. \nonumber \\ && \left.
-M^2\right]\ , 
\label{eq:7.16a}
\end{eqnarray}
for scalar ($s$) and vector ($v$) coupling respectively, $M_B$ is the mass of the exchanged baryon. The $J^P=\frac{1}{2}^-$ baryon direct pole gives rise to the relativistic amplitudes
\begin{eqnarray}
 A_{s}(s,t,u) &=& \frac{g^{*(s)}_{12} g^{*(s)}_{34}}{s- M_B^2 + i\epsilon}\left[M_B + M \right] , \nonumber \\
 B_{s}(s,t,u) &=& \frac{g^{*(s)}_{12} g^{*(s)}_{34}}{s- M_B^2 + i\epsilon}\ ,
\nonumber \\
 A_{v}(s,t,u) &=& \frac{f^{*(v)}_{12} f^{*(v)}_{34}/m_{\pi^+}^2}{s-M_B^2+i\epsilon}\left[\vphantom{M^2M_B} s\left(-M+M_B\right)
\right. \nonumber \\ && \left.
+M^3 -M^2M_B\right]\ , \nonumber \\
B_{v}(s,t,u) &=& -\frac{f^{*(v)}_{12} f^{*(v)}_{34}/m_{\pi^+}^2}{s-M_B^2+i\epsilon}\left[-s+2MM_B
\right. \nonumber \\ && \left.
-M^2\right]\ ,
\label{eq:7.16b}
\end{eqnarray}
for scalar and vector coupling respectively.
The $J^P=\frac{3}{2}^+$ resonance-exchange relativistic amplitudes are more complicated

\begin{eqnarray}
 A_{Y^*}(s,t,u) &=& \frac{f^{*}_{14} f^{*}_{23}/m_{\pi^+}^2}{u- M_{Y^*}^2 + i\epsilon}
\left[\frac{\left[t-2m^2\right]}{2}
 \left(M + M_{Y^*}\right) \right. 
\nonumber \\ && 
 +\frac{M_{Y^*}}{3}\left[u-M^2\right]
\nonumber \\ && \left.
 + \frac{1}{6M_{Y^*}}\left[\vphantom{M^3M_{Y^*}} -u^2+2M u \left(M+M_{Y^*}\right)
 \right.\right. \nonumber \\ && \left.\left. 
-2M^3 M_{Y^*} - M^4 + m^4\right]
\right. \nonumber \\ && \left. 
+\frac{1}{6M_{Y^*}^2}\left[M^2-m^2-u\right]^2
 \left(M+M_{Y^*}\right) \vphantom{\frac{\left[2m^2\right]}{2}}\right] , \nonumber \\ 
 B_{Y^*}(s,t,u) &=& \frac{f^{*}_{14} f^{*}_{23}/m_{\pi^+}^2}{u- M_{Y^*}^2 + i\epsilon}
\left[-\frac{\left[t-2m^2\right]}{2}
 \right. \nonumber \\ && \left. 
 +\frac{1}{3M_{Y^*}}\left[\vphantom{M_{Y^*}M^3} \left(M+M_{Y^*}\right)\left(2M M_{Y^*} -m^2\right)
\right.\right. \nonumber \\ && \left. \left.
- u M + M^3 \vphantom{M_{Y^*}M^3}\right]
 \right. \nonumber \\ 
 && \left. -\frac{1}{6M_{Y^*}^2}\left[u-M^2+m^2\right]^2 \vphantom{\frac{\left[2m^2\right]}{2}}\right]\ ,
\label{eq:7.15} \end{eqnarray}
where $M_{Y^*}$ is the mass of the exchanged resonance. The $J^P=\frac{3}{2}^+$ resonance direct pole gives rise to the relativistic amplitudes
\begin{eqnarray}
 A_{Y^*}(s,t,u) &=& \frac{f^{*}_{12} f^{*}_{34}/m_{\pi^+}^2}{s- M_{Y^*}^2 + i\epsilon}
\left[\frac{\left[t-2m^2\right]}{2}
 \left(M + M_{Y^*}\right) \right. 
\nonumber \\ && 
 +\frac{M_{Y^*}}{3}\left[s-M^2\right] 
\nonumber \\ &&
+ \frac{1}{6M_{Y^*}}\left[\vphantom{M_{Y^*}M^3}-s^2+2M s \left(M+M_{Y^*}\right)
 \right. \nonumber \\ && \left. 
-2M^3 M_{Y^*} - M^4 + m^4\right]
 \nonumber \\ && \left. +\frac{1}{6M_{Y^*}^2}\left[M^2-m^2-s\right]^2
 \left(M+M_{Y^*}\right) \vphantom{\frac{\left[2m^2\right]}{2}}\right]\ , \nonumber \\ 
 B_{Y^*}(s,t,u) &=& -\frac{f^{*}_{12} f^{*}_{34}/m_{\pi^+}^2}{s- M_{Y^*}^2 + i\epsilon}
\left[-\frac{\left[t-2m^2\right]}{2}
 \right. \nonumber \\ && \left. 
 +\frac{1}{3M_{Y^*}}\left[\vphantom{\left(M_{Y^*}m^2\right)M^3}\left(M+M_{Y^*}\right)\left(2M M_{Y^*} -m^2\right) 
\right.\right. \nonumber \\ && \left. \left.
- s M + M^3 \vphantom{\left(M_{Y^*}m^2\right)M^3}\right]
 \right. \nonumber \\ 
 && \left. -\frac{1}{6M_{Y^*}^2}\left[s-M^2+m^2\right]^2 \vphantom{\frac{\left[2m^2\right]}{2}}\right]\ .
\label{eq:7.17} \end{eqnarray}

\subsubsection{Meson- and Pomeron-exchange amplitudes}
The relativistic amplitudes for the $t$-channel Pomeron-exchange, scalar-meson-exchange, vector-meson-exchange and tensor-meson-exchange are

\begin{eqnarray}
A_P(s,t,u) &=& \frac{g_{PPP}\ g_P}{{\cal M}}\ \ ,\ \ B_P(s,t,u) = 0\ , \nonumber \\
 A_S(s,t,u) &=& \frac{g_{PPS}\ g_S}{t- m_S^2 + i\epsilon}\ \ ,\ \  B_S(s,t,u) = 0\ , 
\nonumber \\
 A_V(s,t,u) &=& \frac{g_{PPV}}{t- m_V^2 + i\epsilon}
\frac{f_V}{2{\cal M}}\left[ s - u\right]\ , \nonumber \\
 B_V(s,t,u) &=& -2 \frac{g_{PPV}}{t- m_V^2 + i\epsilon}
\left[g_V + \frac{M}{\cal M}\ f_V\right]\ ,
\nonumber \\
 A_T(s,t,u) &=& \frac{g_{PPT}/m_{\pi^+}}{t- m_T^2 + i\epsilon}
\left[\frac{1}{4}\left( s - u\right)^2 F_2 
-\frac{1}{6}\left[4m^2-t\right]\ 
\right. \nonumber \\ && \left.  \times
 \left[2M F_1 + \frac{1}{2}\left(4M^2-t\right) F_2\right] \vphantom{\frac{1}{4}\frac{1}{6}}\right]\ , \nonumber \\
 B_T(s,t,u) &=& \frac{g_{PPT}/m_{\pi^+}}{t- m_T^2 + i\epsilon}
 \left[ s - u\right]\ F_1\ ,
\label{eq:7.14} \end{eqnarray}
where $m_S$, $m_V$ and $m_T$ are the masses of the exchanged scalar-meson, vector-meson and tensor-meson respectively.

Here we notice that for the meson-exchange and Pomeron-exchange amplitudes an extra factor 2 must be added to the amplitudes if both the initial and final state contain a $\pi$ or $\eta$, this is not the case for any other combination of pseudoscalar-mesons in the initial and final state. For elastic $\pi N$ scattering for example, an extra factor 2 is added to the $\rho$-exchange, Pomeron-exchange and $\sigma$-exchange amplitudes.
\subsection{Partial wave potentials}
\label{sec:8}
As discussed in Sec. \ref{chap:3} we solve the integral equation for the $T$-matrix on the partial wave basis, Eq. (\ref{eq:6.7}). And in paper II
 we fit the NSC model to the $\pi N$ partial wave analysis \cite{Arn95} and the $K^+ N$ partial wave analysis \cite{Hys92}. For this purpose we need to calculate the partial wave projection of the potentials, Eq. (\ref{eq:5.17a}). In our approximation, the potentials are given by the invariant amplitudes $A$ and $B$, Eqs. (\ref{eq:7.10})--(\ref{eq:7.14}), of the one-meson-exchange and one-baryon-exchange Feynman diagrams.

Until this point, we did not mention the need for form factors to regulate the high energy behavior, i.e. the short distance behavior, of the potentials, but in fact the kernel of the integral equation without form factors does not satisfy the Fredholm condition, $\int\int \ dp\ dk\ \left|K(p,k)\right|^2<\infty$, in general.
Furthermore we have derived our one-meson-exchange and one-baryon-exchange potentials from Quantum Field Theory, which is in principle only valid for point particles, while mesons and baryons have an internal structure. Therefore we need  to take into account the extended size of the mesons and baryons by means of a form factor. Since the ground state wave functions of the quarks are Gaussian, form factors of the Gaussian type are used in the NSC model. For $t$-channel exchanges we multiply the potentials by the form factor
\begin{equation}
F(\Lambda )=e^{-({\bf p}_f-{\bf p}_i)^2/\Lambda^2}\ ,
\end{equation}
where ${\bf p}_i$ and ${\bf p}_f$ are the CM three-momenta for the initial and final state respectively, i.e. at both vertices we have used the difference between the final and initial three-momenta. $\Lambda$ is a cutoff mass, which will be determined in the fit to the experimental phases.

For $u$- and $s$-channel exchanges, the difference between the final and initial three-momenta of the baryon is used, giving the form factor
\begin{equation}
F(\Lambda )=e^{-({\bf p}_f^2+{\bf p}_i^2)/2\Lambda^2}\ .
\end{equation}
This form factor obviously does not depend on the scattering angle $\theta$, which makes the partial wave projection easier. For the $u$- and $t$-channel we rewrite the denominators of the potentials in the form
\begin{eqnarray}
\frac{1}{t-m^2}&=&\frac{-1}{2p_fp_i}\frac{1}{z_t-x}\ , \nonumber \\
\frac{1}{u-m^2}&=&\frac{-1}{2p_fp_i}\frac{1}{z_u+x}\ ,
\end{eqnarray}
where $x=\cos (\theta )$ and $\theta $ is the angle between the final and initial three-momenta ${\bf p}_f$ and ${\bf p}_i$. Here we have defined the $z_t$ and $z_u$ factors
\begin{eqnarray}
z_t&=&\frac{1}{2p_fp_i}\left[m^2+p_f^2+p_i^2-\frac{1}{4}\left[E_i-E_f-\omega _i+\omega _f\right]^2\right]\, , \nonumber \\
z_u&=&\frac{1}{2p_fp_i}\left[m^2+p_f^2+p_i^2-\frac{1}{4}\left[E_i+E_f-\omega _i-\omega _f\right]^2\right]\, ,
\nonumber \\
\label{eq:5.32}
\end{eqnarray}
where $E_{f,i}$ are the baryon energies, $\omega _{f,i}$ are the meson energies and $m$ the mass of the exchanged particle. For positive and real momenta, i.e. for open channels, we have $z>1$. Now it is clear that the potentials $V^{(\alpha )}$ of Eq. (\ref{eq:3.25}), where $\alpha $ stands for central or spin-orbit, can be expanded in $x$ as
\footnote{In case of more complicated exchanges, e.g. $J^{PC}=\frac{3}{2}^+$-resonance, the expansions of the potentials have an additional term of higher order in $x$, for the $t$- and $u$-channel $x^3U^{(\alpha )}$, and for the $s$-channel $x^2Z^{(\alpha )}$.\\
We notice that a similar expansion for $F$ and $G$ instead of $V^{(C)}$ and $V^{(SO)}$ would be a little simpler. However we will use the central and spin-orbit potentials in light of a momentum space version of the NSC model.}
\begin{eqnarray}
V^{(\alpha )}({\bf p}_f,{\bf p}_i)&=&\frac{1}{2p_fp_i}\left[X^{(\alpha )}+xY^{(\alpha )}+x^2Z^{(\alpha )}\right]\frac{F(\Lambda _t)}{z_t-x}\ ,\nonumber \\
V^{(\alpha )}({\bf p}_f,{\bf p}_i)&=&\frac{1}{2p_fp_i}\left[X^{(\alpha )}+xY^{(\alpha )}+x^2Z^{(\alpha )}\right]\frac{F(\Lambda _u)}{z_u+x}\ ,\nonumber \\
V^{(\alpha )}({\bf p}_f,{\bf p}_i)&=&\left[X^{(\alpha )}+xY^{(\alpha )}\right]\frac{F(\Lambda _s)}{s-M_B^2}\ ,
\end{eqnarray}
for $t$-, $u$- and $s$-channel exchanges respectively, for all particles that are exchanged. The coefficients $X^{(\alpha )}$, $Y^{(\alpha )}$ and $Z^{(\alpha )}$ can be found easily by writing out the $x$-dependence of the invariant amplitudes $A$ and $B$, they are listed in Appendix \ref{app:I}  for each type of exchange. 

The partial wave potentials $V^{(\alpha )}_L$ are found by inverting the partial wave expansion Eq. (\ref{eq:b5.6}), giving
\begin{equation}
V^{(\alpha )}_L(p_f,p_i)=\frac{1}{2}\int_{-1}^1dxP_L(x)V^{(\alpha )}({\bf p}_f,{\bf p}_i)\ .
\end{equation}
 The partial wave potentials now take the form
\begin{eqnarray}
\lefteqn{
V^{(\alpha )}_L(p_f,p_i)=\left[X^{(\alpha)}\frac{F(\Lambda_s)}{s-M_B^2}\delta_{L,0}+Y^{(\alpha)}\frac{F(\Lambda_s)}{s-M_B^2}\frac{\delta_{L,1}}{3}\right] \nonumber 
} \\
\lefteqn{
V^{(\alpha )}_L(p_f,p_i)=\frac{1}{2p_fp_i}\left[\left(X^{(\alpha )}+z_tY^{(\alpha )}+z^2_tZ^{(\alpha )}\right)
\right.
}
\nonumber \\ &&
\times U_L(\Lambda _t,z_t)-\left(Y^{(\alpha )}+z_tZ^{(\alpha )}\right)R_L(\Lambda _t,z_t)
\nonumber \\ && \left.-Z^{(\alpha )}S_L(\Lambda _t,z_t)\right]\ ,
\nonumber \\
\lefteqn{
V^{(\alpha )}_L(p_f,p_i)=\frac{(-1)^L}{2p_fp_i}\left[\left(X^{(\alpha )}-z_uY^{(\alpha )}+z^2_uZ^{(\alpha )}\right)
\right.
}
\nonumber \\ &&
\times U_L(\Lambda _u,z_u)-\left(-Y^{(\alpha )}+z_uZ^{(\alpha )}\right)R_L(\Lambda _u,z_u)
\nonumber \\ && \left.-Z^{(\alpha )}S_L(\Lambda _u,z_u)\right]\ , 
\label{eq:7.60}
\end{eqnarray}
for $s$-, $t$-, and $u$-channel exchanges respectively. We have defined the basic partial wave projections $U_L$, $R_L$, $S_L$ and $T_L$ in terms of the Legendre polynomials $P_L(x)$ and the form factors
\begin{eqnarray}
U_L(\Lambda ,z)&=&\frac{1}{2} \int_{-1}^1 dx\frac{P_L(x)F(\Lambda )}{z-x}\ , \nonumber \\
R_L(\Lambda ,z)&=&\frac{1}{2} \int_{-1}^1 dxP_L(x)F(\Lambda )\ , \nonumber \\
S_L(\Lambda ,z)&=&\frac{1}{2} \int_{-1}^1 dxP_L(x)xF(\Lambda )\ , \nonumber \\
T_L(\Lambda ,z)&=&\frac{1}{2} \int_{-1}^1 dxP_L(x)x^2F(\Lambda )\ .
\label{eq:5.31}
\end{eqnarray}
The factor $(-)^L$ appearing in the $u$-channel partial wave potentials, which is the result of changing the integration variable $x \rightarrow -x$ in the Legendre polynomial, is typical for exchange forces. In this way it can be seen that the total partial wave potential is a linear combination of a direct and an exchange potential, $V^{(\pm)}=V_d\pm V_e$ and the corresponding $T$-matrix is $T^{(\pm)}=T_d\pm T_e$. The amplitudes $T_d$ and $T_e$ do not satisfy an integral equation, but the two linear combinations $T^{(\pm)}=T_d\pm T_e$ do.

We notice that if the form factor does not depend on x (in case of $u$-channel potentials) or if we consider the limit $\Lambda \rightarrow \infty$, i.e. $F(\Lambda ) \rightarrow 1$, the basic partial wave projections defined in Eq. (\ref{eq:5.31}) are proportional to the simple functions
\begin{eqnarray}
U_L(\Lambda ,z)&\propto&Q_L(z)\ , \nonumber \\
R_L(\Lambda ,z)&\propto&\delta_{L,0}\ , \nonumber \\
S_L(\Lambda ,z)&\propto&\frac{1}{3}\delta_{L,1}\ , \nonumber \\
T_L(\Lambda ,z)&\propto&\frac{1}{3}\delta_{L,0}+\frac{2}{15}\delta_{L,2}\ ,
\end{eqnarray}
where $Q_L(z)$ is the Legendre function of the second kind, which is an analytic function of its argument except for a cut on the real axis running from --1 to 1, as is clear from Eq. (\ref{eq:5.31}). In view of Eq. (\ref{eq:5.32}), the cut is entered only for on-energy-shell potentials below threshold, but we always calculate the on-energy-shell potentials above threshold, so we will never reach the cut.

In the NSC $\pi N$-model we will include $s$-channel baryon-exchange diagrams, which are in principle separable diagrams, having the form
\begin{equation}
V\left(p_f,p_i\right)=\frac{\Gamma \left(p_f\right)\ \Gamma \left(p_i\right)}{\sqrt{s}\pm M_0}\ .
\end{equation}
Writing out the partial wave potential for the $\Delta$ pole ($P_{33}$-wave) explicitly, using  Eq. (\ref{eq:5.17a}) and Eq. (\ref{eq:7.60}),
\begin{eqnarray}
V_{33}&=&\frac{f_{\pi N\Delta}^2}{m_{\pi ^+}^2}\ \frac{1}{3} \  \sqrt{\left(E_i+M_i\right)\left(E_f+M_f\right)}
 \nonumber \\ && \times
\ p_fp_i\ \frac{1}{\sqrt{s}-M_0}\ ,
\label{eq:7.30}
\end{eqnarray} 
we see that this potential is of the separable kind indeed.



We need to be careful in including the $s$-channel diagrams in a model that has been renormalized, i.e. in which (renormalized) physical coupling constants and masses are used. It is not possible to simply add the $s$-channel diagrams to the other ones, because iterations of $s$-channel diagrams will give contributions to the vertex and self-energy.
The way these diagrams are included in the NSC model is described in paper II,
here we show that bare masses and coupling constants should be used in the $s$-channel diagrams and that these bare parameters are determined by requiring that (i) the $T$-matrix has a pole at the physical mass $\sqrt{s}=M_e$, (ii) the residue at the pole is given by the physical coupling constant.
\section{Summary}
Analogous to the Nijmegen soft-core one-boson-exchange $NN$ and $YN$ models, we have derived the NSC model for the interaction between pseudoscalar-mesons and baryons ($\pi N$, $K^+ N$, etc.).

For the general mass case the meson-baryon potentials in the context of a relativistic two-particle equation, the Bethe-Salpeter equation, are defined in Sec. \ref{chap:2}. The potentials consist of one-meson and one-baryon-exchange Feynman diagrams. The Bethe-Salpeter equation is approximated by assuming dynamical pair-suppression¨, hence neglecting the propagation of negative-energy states, and by integrating over the time variable, giving a three-dimensional integral equation for the scattering amplitude, which is a generalization of the Lippmann-Schwinger equation.

A transformation of this equation on the plane wave basis to the partial wave (LSJ) basis is described in Sec. \ref{chap:3}. A one-dimensional integral equation for the partial wave scattering amplitude is derived, which is decoupled for each partial wave, because of parity conservation in strong interactions.

In Sec. \ref{chap:5} the interaction Hamiltonians are given and the resulting one-baryon-exchange and one-meson-exchange invariant amplitudes have been derived, these amplitudes define the partial wave potentials used in the calculations.
\begin{acknowledgements}
The authors would like to thank Prof. J. J. de Swart and Prof. R. G. E. Timmermans for stimulating discussions.
\end{acknowledgements}
\appendix
\section{Matrix elements on the LSJ-basis}
\label{app:AAA}
\subsection{Partial wave amplitudes}
\label{app:bB}
\begin{widetext}
Here we derive the central and non-central potential matrix elements on the LSJ-basis in Eqs. (\ref{eq:5.4}) and (\ref{eq:5.9})\\ \\
\underline{
\noindent 1.\ \textit{Central} $P_1=1$:
}
\begin{eqnarray}
&& \left(q_f;L'J'M'|F({\bf q}_f,{\bf q}_i)|q_i;LJM\right) =
\sum_{s_f,s_i}\int \frac{d^3q'_f}{(2\pi)^3}\ \int \frac{d^3q'_i}{(2\pi)^3}\ 
\nonumber \\ && \hspace{.5cm} \times
\left(q_f;L'J'M'|{\bf q'}_f,s_f\right) 
 \left({\bf q'}_f,s_f|F_{op}|{\bf q'}_i,s_i\right)\ 
\left({\bf q'}_i,s_i|q_i,LJM\right) .           
\label{eq:b5.4} \end{eqnarray}
We now use the matrix elements
\begin{eqnarray}
 \left({\bf q'}_f,s_f|F_{op}|{\bf q'}_i,s_i\right) &=&
F({\bf q'}_f,{\bf q'}_i) \delta_{s_f,s_i} 
=
4\pi\sum_{l=0}^\infty \ F_l(q'_f,q'_i)
\sum_{n=-l}^{l} \ Y^l_n(\hat{\bf q'}_f)
 Y^l_n(\hat{\bf q'}_i)^*\ \delta_{s_f,s_i}\ ,
\nonumber \\
\left({\bf q'},s|q,LJM\right) &=& (2\pi)^3\ \frac{\delta(q'-q)}{q^2}\ 
{\cal Y}_{JL}^{M}(\hat{\bf q'},s)\ .
\label{eq:b5.6} 
\end{eqnarray}
Then, substituting Eq. (\ref{eq:b5.6}) into Eq. (\ref{eq:b5.4})
and performing the momentum and angular integrals and summations, we find
\begin{eqnarray}
 \left(q_f;L'J'M'|F({\bf q}_f,{\bf q}_i)|q_i;LJM\right) &=&
4\pi F_L\left(q_f,q_i\right)\ 
 \delta_{L',L} \delta_{J',J}\ \delta_{M',M}\ .
\label{eq:b5.8} \end{eqnarray}
\\ 
\underline{
\noindent 2.\ \textit{ Non-central} 
  $P'_{2}= (\mbox{\boldmath $\sigma$}\cdot\hat{\bf q}_f)
  (\mbox{\boldmath $\sigma$}\cdot\hat{\bf q}_i)$:
}
\begin{eqnarray}
&& \left(q_f;L'J'M'|G({\bf q}_f,{\bf q}_i)|q_i;LJM\right) =
\sum_{s_f,s_i}\int \frac{d^3q'_f}{(2\pi)^3}\ \int \frac{d^3q'_i}{(2\pi)^3}\ 
\nonumber \\ && \hspace{.5cm} \times
\left(q'_f;L'J'M'|{\bf q'}_f,s_f\right) 
 \left({\bf q'}_f,s_f|G_{op}|{\bf q'}_i,s_i\right)\ 
\left({\bf q'}_i,s_i|q_i,LJM\right)\ ,           
\label{eq:b5.9} \end{eqnarray}
where analogously to Eq. (\ref{eq:b5.6}),
\begin{eqnarray}
 \left({\bf q'}_f,s_f|G_{op}|{\bf q'}_i,s_i\right) =
4\pi                       
  \sum_{l=0}^\infty  G_l(q'_f,q'_i)
\sum_{n=-l}^{l}\ Y^l_n(\hat{\bf q'}_f) Y^l_n(\hat{\bf q'}_i)^*
\cdot
\sum_{s}\ \left( s_f| (\mbox{\boldmath $\sigma$}\cdot\hat{\bf q}_f')|s\right)\ \left( s|(\mbox{\boldmath $\sigma$}\cdot\hat{\bf q}_i')|s_i\right)
\ .
\label{eq:b5.11} \end{eqnarray}
Using Eq. (\ref{eq:b5.6}) and substituting Eq. (\ref{eq:b5.11}) into Eq. (\ref{eq:b5.9}), and performing the momentum 
integrals, we find
\begin{eqnarray}
\left(q_f;L'J'M'|G_{op}|q_i;LJM\right) &=& 
  4\pi \sum_{s_f,s_i,s}\ \sum_{l}\ 
 G_l\left(q_f,q_i\right)\sum_{n=-l}^{l} \nonumber \\
 && 
\times
\int\! d\hat{\Omega}_{q_f}\ {\cal Y}^{M'}_{J'L'}(\hat{\bf q}_f,s_f)^* 
(s_f|\left(\mbox{\boldmath $\sigma$}\cdot\hat{\bf q}_f\right)|s)\ Y^l_n (\hat{\bf q}_f) 
\nonumber \\ 
&& 
\times 
 \int\! d\hat{\Omega}_{q_i}\ Y^l_n(\hat{\bf q}_i)^* 
(s|\left(\mbox{\boldmath $\sigma$}\cdot\hat{\bf q}_i\right)|s_i)\  
{\cal Y}^{M}_{JL}(\hat{\bf q}_i,s_i)
\ .
\label{eq:b5.12} \end{eqnarray}
In the two-dimensional $L=J\mp \frac{1}{2}$-space, the 
$(\mbox{\boldmath $\sigma$}\cdot\hat{\bf q})$-operator has the matrix elements,
see Appendix \ref{app:B},
\begin{equation}
 \sum_{s'} (s|\left(\mbox{\boldmath $\sigma$}\cdot\hat{\bf q}\right)|s')\  
 {\cal Y}^{M}_{JL}(\hat{\bf q},s') = \sum_{L'}
 {\cal Y}^{M}_{JL'}(\hat{\bf q},s)\ a_{L',L}\ \ {\rm where}\ 
 a_{L',L} = \left(\begin{array}{cc} 0 & -1 \\ -1 & 0 \end{array}\right)\ .
\label{eq:5.13} \end{equation}
The angular integrals in Eq. (\ref{eq:b5.12}) can now be performed easily.
Then, the result for the non-central amplitude is
\begin{eqnarray}
\left(q_f;L'J'M'|G|q_i;LJM\right) &=&
4\pi \sum_{L''}\ 
 a_{L',L''} G_{L''}\left(q_f,q_i\right)\ a_{L'',L}
 \delta_{J',J}\delta_{M',M}\ .
\label{eq:b5.15} \end{eqnarray}
\subsection{LSJ representation operator}
\label{app:B}
In this appendix we derive Eq. (\ref{eq:5.13}). The spherical wave functions in momentum space with quantum numbers $J$, $L$, $S$, are for spin-0 spin-1/2 given by \cite{SYM}
\begin{equation}
  {\cal Y}_{J L}^{M}(\hat{\bf p},s)= \sum_{m,\mu}
  \mbox{\large\bf $C$}^{L\  \frac{1}{2}\ \: J}_{m\ \mu \ M}\ 
  Y^{L}_{m}(\hat{\bf p})\ \chi^{\left(\frac{1}{2}\right)}_{\mu}(s)\ ,
\label{AppB.1}
 \end{equation}
where $\chi$ is the baryon spin wave function. 
Using the definition for ${\cal Y}^M_{J L}$, Eq. (\ref{AppB.1}), we have 
\begin{eqnarray}
 \sum_{s} (s'|\left(\mbox{\boldmath $\sigma$}\cdot\hat{\bf p}\right)|s)\  
 {\cal Y}^{M}_{J L}(\hat{\bf p},s) &=& 
 (-)^m \hat{\bf p}_m\ (s'|\sigma_{-m}|s)\  
 C^{L\ \ \frac{1}{2}\ \, J}_{m_l\ \mu\ M}\ Y^L_{m_l}(\hat{\bf p})\ \chi^{\left(\frac{1}{2}\right)}_\mu(s)
 \nonumber \\ &=& (-)^m \hat{\bf p}_m\ (s'|\sigma_{-m}|s)\  
 C^{L\ \ \frac{1}{2}\  J}_{m_l\ s\ M}\ Y^L_{m_l}(\hat{\bf p})\ ,
\label{AppB.3} \end{eqnarray}
where we used the convention of summation over repeated indices, and 
quantization along the z-axis, which defines the spin variables $s,s'$.
Now, we use the expressions
\begin{eqnarray}
 \hat{\bf p}_m\ Y^L_{m_l}(\hat{\bf p}) &=& \sqrt{\frac{4\pi}{3}}
 Y^1_m(\hat{\bf p})\ Y^L_{m_l}(\hat{\bf p}) \nonumber \\
&=&  \sqrt{\frac{4\pi}{3}}\left[\frac{3(2L+1)}{4\pi(2L'+1)}\right]^{1/2}
 C^{L\ \: 1\ L'}_{\, 0\ \ 0\ 0}\
 C^{L\ \ \, 1\ \: L'}_{m_l\ m\ m'_l}\ Y^{L'}_{m'_l}(\hat{\bf p})\ , \nonumber \\
  (s'|\sigma_{-m}|s) &=& \sqrt{3}\ C^{\, \frac{1}{2}\ \ \: 1\ \; \frac{1}{2}}_{\ s\ -m\ s'}\ ,\nonumber \\
(-)^m &=& -\sqrt{3}\ C^{1\ \ \ \: 1\ \ 0}_{m\ -m\ 0}\ .
\label{AppB.4} \end{eqnarray}
{}From the definition of the $9j$-coefficient \cite{Edm57}, 
these formulas give for Eq. (\ref{AppB.3}) the result
\begin{eqnarray}
 \sum_{s} (s'|\left(\mbox{\boldmath $\sigma$}\cdot\hat{\bf p}\right)|s)\  
 {\cal Y}^{M}_{J L}(\hat{\bf p},s) &=&               
 -3 \sum_{L'}\ \left[\frac{(2L+1)}{(2L'+1)}\right]^{1/2}\ 
C^{L\ \: 1\ L'}_{\, 0\ \ 0\ 0}
\left[\begin{array}{ccc} L & \frac{1}{2} & J \\ 1 & 1 & 0 \\ 
 L' & \frac{1}{2} & J \end{array}\right]\
{\cal Y}^M_{J L'}(\hat{\bf p},s')\ .       
\label{AppB.5} \end{eqnarray}
%
%
Evaluating Eq. (\ref{AppB.5}), one finds for the matrix $a$ in Eq. (\ref{eq:5.13}), 
$
 a = \left(\begin{array}{cc} 0 & -1 \\ -1 & 0 \end{array}\right)\ .
$
%
%
\end{widetext}

\section{Relativistic invariant amplitudes}
\label{app:C}
In this appendix the contributions from the various Feynman diagrams to the relativistic invariant amplitudes $A_{fi}(s,t,u)$ and $ B_{fi}(s,t,u)$, defined in Eq. (\ref{eq:2.5}), are given. The results are valid for elastic as well as inelastic reactions. For details of the derivation we refer to \cite{Pol04}.
%
\subsection{Momentum space 
baryon-exchange diagrams}
\begin{center}
  {\bf (i) $J^P=\frac{1}{2}^+$ baryon-exchange} \\
\end{center}
\label{app:D}     
\noindent (i) \underline{pseudoscalar coupling}:\\

\begin{eqnarray}
 A_{ps} &=&
 -\frac{g_{14} g_{23}}{u-M_B^2+i\epsilon}\
 \left[-\frac{M_f+M_i}{2}+ M_B\right]\ , \nonumber \\ && \nonumber \\
 B_{ps} &=&
 -\frac{g_{14} g_{23}}{u-M_B^2+i\epsilon}\ .
\label{appD.4} \end{eqnarray}

\noindent (ii) \underline{pseudovector coupling}: \\
\begin{eqnarray}
 A_{pv} &=& -\frac{f_{14} f_{23}/m_{\pi^+}^2}{u-M_B^2+i\epsilon}\left[\vphantom{\frac{M_f^2}{2}} u\left(\frac{M_f+M_i}{2}+M_B\right)\right.\nonumber \\
&&\left.-\frac{M_f+M_i}{2}M_fM_i -\frac{M_f^2+M_i^2}{2}M_B\right]\ ,\nonumber \\
B_{pv} &=& -\frac{f_{14} f_{23}/m_{\pi^+}^2}{u-M_B^2+i\epsilon}\left[\vphantom{M_f} u+\left(M_f+M_i\right)M_B
\right. \nonumber \\ &&\left.
+M_fM_i\right]\ . 
\end{eqnarray}

\begin{center}
  {\bf (ii) $J^P=\frac{1}{2}^+$ pole diagram} \\
\end{center}
Using crossing symmetry \cite{Gas66}, we can cross the results of the $u$-channel baryon-exchange into the $s$-channel and obtain the invariant amplitudes $A_{fi}(s,t,u)$ and $B_{fi}(s,t,u)$ for the pole diagram. We have to replace $q\rightarrow -q'$ and $q'\rightarrow -q$, which means that we have to make the substitutions $u\leftrightarrow s$, $ m_f^2 \leftrightarrow m_i^2$ and add a minus sign to the amplitude $B$ because of Eq. (\ref{eq:2.5}). The $J^P=\frac{1}{2}^+$ pole amplitudes are\\ \\

\noindent (i) \underline{pseudoscalar coupling}: 
\begin{eqnarray}
 A_{ps} &=&
 -\frac{g_{12} g_{34}}{s-M_B^2+i\epsilon}\
 \left[-\frac{M_f+M_i}{2}+ M_B\right]\ , \nonumber \\ && \nonumber \\
 B_{ps} &=&
 \frac{g_{12} g_{34}}{s-M_B^2+i\epsilon}\ .
\label{appD.4a} \end{eqnarray}

\noindent (ii) \underline{pseudovector coupling}: 

\begin{eqnarray}
 A_{pv} &=& -\frac{f_{12} f_{34}/m_{\pi^+}^2}{s-M_B^2+i\epsilon}\left[\vphantom{\frac{M_f^2}{2}} s\left(\frac{M_f+M_i}{2}+M_B\right)\right.\nonumber \\
&&\left.-\frac{M_f+M_i}{2}M_fM_i -\frac{M_f^2+M_i^2}{2}M_B\right]\ ,\nonumber \\
B_{pv} &=& \frac{f_{12} f_{34}/m_{\pi^+}^2}{s-M_B^2+i\epsilon}\left[s+\left(M_f+M_i\right)M_B
\right. \nonumber \\ &&\left.
+M_fM_i\right]\ . 
\end{eqnarray}

\begin{center}
  {\bf (iii) $J^P=\frac{1}{2}^-$ baryon-exchange} \\
\end{center}
\noindent (i) \underline{scalar coupling}:\\

\begin{eqnarray}
 A_{s} &=&
 -\frac{g^{*(s)}_{14} g^{*(s)}_{23}}{u-M_B^2+i\epsilon}\
 \left[-\frac{M_f+M_i}{2}- M_B\right]\ , \nonumber \\ && \nonumber \\
 B_{s} &=&
 -\frac{g^{*(s)}_{14} g^{*(s)}_{23}}{u-M_B^2+i\epsilon}\ .
\label{appD.4m} \end{eqnarray}

\noindent (ii) \underline{vector coupling}: \\

\begin{eqnarray}
 A_{v} &=& -\frac{f^{*(v)}_{14} f^{*(v)}_{23}/m_{\pi^+}^2}{u-M_B^2+i\epsilon}\left[\vphantom{\frac{M_f^2}{2}} u\left(\frac{M_f+M_i}{2}-M_B\right)\right.\nonumber \\
&&\left.-\frac{M_f+M_i}{2}M_fM_i +\frac{M_f^2+M_i^2}{2}M_B\right]\ ,\nonumber \\
B_{v} &=& -\frac{f^{*(v)}_{14} f^{*(v)}_{23}/m_{\pi^+}^2}{u-M_B^2+i\epsilon}\left[u-\left(M_f+M_i\right)M_B
\right. \nonumber \\ && \left.
+M_fM_i\right]\ . 
\end{eqnarray}

\begin{center}
  {\bf (iii) $J^P=\frac{1}{2}^-$ pole diagram} \\
\end{center}
Applying crossing symmetry again we find, similar to the $J^P=\frac{1}{2}^+$ baryon pole diagram, the invariant amplitudes for the $J^P=\frac{1}{2}^-$ baryon pole diagram.\\

\noindent (i) \underline{scalar coupling}:

\begin{eqnarray}
 A_{s} &=&
 -\frac{g^{*(s)}_{12} g^{*(s)}_{34}}{s-M_B^2+i\epsilon}\
 \left[-\frac{M_f+M_i}{2}- M_B\right]\ , \nonumber \\ && \nonumber \\
 B_{s} &=&
 \frac{g^{*(s)}_{12} g^{*(s)}_{34}}{s-M_B^2+i\epsilon}\ .
\label{appD.4am} \end{eqnarray}

\noindent (ii) \underline{vector coupling}: 

\begin{eqnarray}
 A_{v} &=& -\frac{f^{*(v)}_{12} f^{*(v)}_{34}/m_{\pi^+}^2}{s-M_B^2+i\epsilon}\left[\vphantom{\frac{M_f^2}{2}} s\left(\frac{M_f+M_i}{2}-M_B\right)\right.\nonumber \\
&&\left.-\frac{M_f+M_i}{2}M_fM_i +\frac{M_f^2+M_i^2}{2}M_B\right]\ ,\nonumber \\
B_{v} &=& \frac{f^{*(v)}_{12} f^{*(v)}_{34}/m_{\pi^+}^2}{s-M_B^2+i\epsilon}\left[s-\left(M_f+M_i\right)M_B
\right. \nonumber \\ && \left.
+M_fM_i\right]\ . 
\end{eqnarray}

\begin{center}
  {\bf (v) $J^P=\frac{3}{2}^+$ baryon-exchange} \\
\end{center}
\label{app:F}     

\begin{eqnarray}
 A_{Y^*} &=& \frac{f^{*}_{14} f^{*}_{23}/m^2_{\pi^+}}{u-M_{Y^*}^2+i\epsilon}
 \left[
\frac{t-m_f^2-m_i^2}{2}
\right. \nonumber \\ && \times
\left[\frac{M_f+M_i}{2}+M_{Y^*}\right]
\nonumber \\ && \left.
 +\frac{1}{6M_{Y^*}^2}\left[M_f^2-m_i^2-u\right]\left[M_i^2-m_f^2-u\right]
\right.  \nonumber \\ && \left. \times
 \frac{M_f+M_i}{2}
 + \frac{M_{Y^*}}{3}\left[u-\frac{M_f^2+M_i^2}{2}\right]
 \right. \nonumber \\ &&
 +\frac{1}{3}\left[\frac{M_f+M_i}{2} u -\frac{\left(M_f^2+M_i^2\right)\left(M_f+M_i\right)}{4}
\right. \nonumber \\ && \left.
+\frac{\left(m_i^2-m_f^2\right)\left(M_f-M_i\right)}{4}\right] 
 +\frac{1}{6M_{Y^*}}
\times \nonumber \\ &&
\left[\left(M_f^2-m_i^2-u\right)\left(\frac{M_i}{2}\left(M_i-M_f\right)-m_f^2\right)-\right. \nonumber \\&&\left.\left.
\left(M_i^2-m_f^2-u\right)\left(\frac{M_f}{2}\left(M_i-M_f\right)+m_i^2\right)\right]
 \vphantom{\frac{m_f^2}{2}}\right]\ , \nonumber
\label{appF.7} 
\end{eqnarray}

\begin{eqnarray}
 B_{Y^*} &=& \frac{f^{*}_{14} f^{*}_{23}/m^2_{\pi^+}}{u-M_{Y^*}^2+i\epsilon}
 \left[-\frac{t-m_f^2-m_i^2}{2} 
\right. \nonumber \\ && \left.
 -\frac{1}{6M_{Y^*}^2}\left[u-M_f^2+m_i^2\right]
\left[u-M_i^2+m_f^2\right] 
 \right. \nonumber \\ &&
+\frac{M_{Y^*}}{3}\left(M_f+M_i\right)-\frac{m_f^2+m_i^2-\left(M_f+M_i\right)^2}{6} \nonumber \\ 
&&\left. 
+\frac{1}{6M_{Y^*}}\left[M_f\left(M_i^2-m_f^2-u\right)
\right. \right. \nonumber \\ && \left. \left.
+M_i\left(M_f^2-m_i^2-u\right)\right]
\vphantom{\frac{m_f^2}{2}}\right]\ .               
\label{appF.8} \end{eqnarray}
 
\begin{center}
  {\bf (vi) $J^P=\frac{3}{2}^+$ pole diagram} \\
\end{center}
\label{app:Fa}
Applying crossing symmetry again we find, similar to the $J^P=\frac{1}{2}^+$ baryon pole diagram, the invariant amplitudes for the $J^P=\frac{3}{2}^+$ baryon pole diagram

\begin{eqnarray}
 A_{Y^*}  &=& \frac{f^{*}_{12} f^{*}_{34}/m^2_{\pi^+}}{s-M_{Y^*}^2+i\epsilon}
 \left[
\frac{t-m_f^2-m_i^2}{2}
\right. \nonumber \\ && \times
\left[\frac{M_f+M_i}{2}+M_{Y^*}\right]
\nonumber \\ && \left.
 +\frac{1}{6M_{Y^*}^2}\left[M_f^2-m_f^2-s\right]\left[M_i^2-m_i^2-s\right]
\right. \nonumber \\ && \left. \times
 \frac{M_f+M_i}{2}
 + \frac{M_{Y^*}}{3}\left[s-\frac{M_f^2+M_i^2}{2}\right]
 \right. \nonumber \\ &&
 +\frac{1}{3}\left[\frac{M_f+M_i}{2} s -\frac{\left(M_f^2+M_i^2\right)\left(M_f+M_i\right)}{4}
\right. \nonumber \\ && \left.
+\frac{\left(m_f^2-m_i^2\right)\left(M_f-M_i\right)}{4}\right] 
 +\frac{1}{6M_{Y^*}}
\times \nonumber \\ &&
\left[\left(M_f^2-m_f^2-s\right)\left(\frac{M_i}{2}\left(M_i-M_f\right)-m_i^2\right)-\right. \nonumber \\&&\left.\left.
\left(M_i^2-m_i^2-s\right)\left(\frac{M_f}{2}\left(M_i-M_f\right)+m_f^2\right)\right]
 \vphantom{\frac{m_f^2}{2}}\right]\ , \nonumber
\label{appFa.7} \end{eqnarray}

\begin{eqnarray}
 B_{Y^*}  &=& -\frac{f^{*}_{12} f^{*}_{34}/m^2_{\pi^+}}{s-M_{Y^*}^2+i\epsilon}
 \left[-\frac{t-m_f^2-m_i^2}{2} 
\right. \nonumber \\ && 
 -\frac{1}{6M_Y^2}\left[s-M_f^2+m_f^2\right]
\left[s-M_i^2+m_i^2\right] 
\nonumber \\ &&
+\frac{M_{Y^*}}{3}\left(M_f+M_i\right)-\frac{m_f^2+m_i^2-\left(M_f+M_i\right)^2}{6} \nonumber \\ 
&&\left. 
+\frac{1}{6M_{Y^*}}\left[M_f\left(M_i^2-m_i^2-s\right)
\right. \right. \nonumber \\ && \left. \left.
+M_i\left(M_f^2-m_f^2-s\right)\right]
\vphantom{\frac{m_f^2}{2}}\right]\ .    
\label{appFa.8} \end{eqnarray}

\subsection{Momentum space meson-exchange diagrams}
\label{app:EE}     
\begin{center}
  {\bf (i) $J^P=0^{++}$ scalar-meson-exchange} \\
\end{center}

\begin{eqnarray}
A_{S}&=&\frac{g_{PPS}\ g_{S}}{t - m^2_S + i\epsilon}\ , \nonumber \\
B_{S}&=&0 \ .
\label{appE.2}
\end{eqnarray}
\vspace{5mm}

\begin{center}
  {\bf (ii) $J^P=1^{--}$ vector-meson-exchange} \\
\end{center}

\begin{eqnarray}
 A_{V} &=& \frac{g_{PPV}}{t-m^2_V+ i\epsilon}
 \left[g_V\ \frac{m_f^2-m_i^2}{m^2_V}
\left(M_i-M_f\right) 
\right. \nonumber \\ && \left.
+ \frac{s-u}{2{\cal M}} f_V \vphantom{\frac{m_f^2}{m_V^2}}\right]\ ,
 \nonumber \\ && \nonumber \\
 B_{V} &=& -2 \frac{g_{PPV}}{t-m^2_V+ i\epsilon}
 \left[f_V \frac{M_f+M_i}{2{\cal M}}
+g_V
\right]\ .
\label{appE.13} \end{eqnarray}
%
%
\begin{center}
  {\bf (iii) $J^P=2^{++}$ tensor-meson-exchange} \\
\end{center}

\begin{eqnarray}
 A_{T} &=& \frac{g_{PPT}/m_{\pi^+}}{\Delta^2-m^2_T+i\epsilon}
 \left[  \left(\frac{s-u}{2}\right)^2\ F_2 
\right.  \nonumber \\ &&
 -\frac{1}{2m^2_T}
\left(m_f^2-m_i^2\right)\left(s-u\right)\left[\left(M_i-M_f\right)\ F_1 
\right. \nonumber \\ && \left.
+ \left(M_i^2-M_f^2\right)\ F_2\right]
 +\frac{1}{2m^4_T}\left(m_f^2-m_i^2\right)^2
\times \nonumber \\ &&
\left(M_i^2-M_f^2\right) \left[\left(M_i-M_f\right) F_1 + \frac{M_i^2-M_f^2}{2} F_2\right]
 \nonumber \\ && 
-\frac{4}{3}\left[-Q^2 + \frac{1}{4m^2_T}\left(m_f^2-m_i^2\right)^2  \right]
\times \nonumber \\ &&
\left[\vphantom{\frac{M_f^2}{m_T^2}} -\frac{1}{2}\left(\left(M_f+M_i\right) F_1 
+ \frac{1}{2}\left(p'+p\right)^2 F_2\right) \right. 
\nonumber \\ && 
\left. + \frac{M_i^2-M_f^2}{2m^2_T} \left(\vphantom{\frac{1}{2}} (M_i-M_f) F_1 
\right. \right. \nonumber \\ && \left. \left.
+ \frac{1}{2}\left(M_i^2-M_f^2\right) F_2\right)
 \right] \left.\vphantom{\left(\frac{s}{2}\right)^2}\right]\ ,
 \nonumber \\ && \nonumber \\
 B_{T} &=& \frac{g_{PPT}/m_{\pi^+}}{\Delta^2-m^2_T+i\epsilon}
 \left[\vphantom{\frac{\left(M_f^2\right)}{m_T^2}}  \left(s-u\right)\ F_1
\right. \nonumber \\ && \left.
  -\frac{\left(m_f^2-m_i^2\right)\left(M_i^2-M_f^2\right)}{m^2_T}\ F_1 \right]\ .
\label{appE.21} \end{eqnarray}

\subsection{Momentum space Pomeron-exchange}
\label{app:DB} 
\begin{eqnarray}
A_{P}&=&\frac{g_{PPP}\ g_{P}}{{\cal M}}\ , \nonumber \\
B_{P}&=&0\ .
\end{eqnarray}   

\section{X,Y,Z-coefficients}
\label{app:I}
Here we list  the explicit expressions for the expansion coefficients $X^{(\alpha )}$, $Y^{(\alpha )}$, $Z^{(\alpha )}$ and $U^{(\alpha )}$ of the partial wave potentials, Eq. (\ref{eq:7.60}), for each type of exchange in the $s$-, $u$- and $t$-channel. We have introduced the notation $N^\pm_{fi}=\sqrt{\left(E_i\pm M_i\right)\left(E_f\pm M_f\right)}$.
\begin{widetext}

\subsection{Baryon-exchange}
\begin{center}
  {\bf (i) $J^P=\frac{1}{2}^+$ baryon-exchange} \\
\end{center}
\label{app:I.3}
(i) \underline{pseudoscalar coupling:}
\begin{eqnarray}
X^{(C)}_B&=&g_{14}g_{23} N^+_{fi}\left[M_B+\frac{W_f+W_i}{2}-M_f-M_i\right]\ , \nonumber \\
Y^{(C)}_B&=&g_{14}g_{23}\ N^-_{fi}\left[-M_B+\frac{W_f+W_i}{2}+M_f+M_i\right]\ , \nonumber \\
X^{(SO)}_B&=&-g_{14}g_{23}\ N^-_{fi}\left[-M_B+\frac{W_f+W_i}{2}+M_f+M_i\right]\ .
\end{eqnarray}

\noindent (ii) \underline{pseudovector coupling:}
\begin{eqnarray}
X^{(C)}_B&=&-\frac{f_{14}f_{23}}{m_{\pi^+}^2}\ N^+_{fi}
\left[\left(-\frac{M_f+M_i}{2}-M_B\right)
\left(\left(\frac{E_f+E_i-\omega_f-\omega_i}{2}\right)^2-p_f^2-p_i^2\right)
+\frac{M_f+M_i}{2}M_fM_i
\right. \nonumber \\ &&
+\frac{M_f^2+M_i^2}{2}M_B
-\frac{W_f+W_i-M_f-M_i}{2}\left( \left(\frac{E_f+E_i-\omega_f-\omega_i}{2}\right)^2
-p_f^2-p_i^2 
\left.+\left(M_f+M_i\right)M_B
\right. \right. \nonumber \\ && \left. \left.
+M_fM_i\vphantom{\left(\left(\frac{E_f}{2}\right)^2\right)} \right) \vphantom{\left(\left(\frac{E_f}{2}\right)^2\right)}\right]\ , 
\nonumber \\
Y^{(C)}_B&=&-\frac{f_{14}f_{23}}{m_{\pi^+}^2}\left[\vphantom{\left(\left(\frac{E_f}{2}\right)^2\right)} N^+_{fi}
\left[-\left(-\frac{M_f+M_i}{2}-M_B\right)
+\frac{W_f+W_i-M_f-M_i}{2}\right]2p_fp_i 
+N^-_{fi}
\left[\vphantom{\left(\left(\frac{E_f}{2}\right)^2\right)}-\left(-\frac{M_f+M_i}{2}
\right. \right. \right. \nonumber \\ && \left. \left.
-M_B\right)\left(\left(\frac{E_f+E_i-\omega_f-\omega_i}{2}\right)^2-p_f^2-p_i^2\right)
-\frac{M_f+M_i}{2}M_fM_i-\frac{M_f^2+M_i^2}{2}M_B
\right. \nonumber  \\&& \left. \left.
-\frac{W_f+W_i+M_f+M_i}{2}
\left(\left(\frac{E_f+E_i-\omega_f-\omega_i}{2}\right)^2-p_f^2-p_i^2
+\left(M_f+M_i\right)M_B+M_fM_i\vphantom{\frac{A}{A}}\right)\right]\right]\ ,  \nonumber \\
Z^{(C)}_B&=&-\frac{f_{14}f_{23}}{m_{\pi^+}^2}\ N^-_{fi} \left[-\frac{M_f+M_i}{2}-M_B+\frac{W_f+W_i+M_f+M_i}{2}\right]2p_fp_i \ , \nonumber \\
X^{(SO)}_B&=&\frac{f_{14}f_{23}}{m_{\pi^+}^2}\ N^-_{fi}
\left[-\left(-\frac{M_f+M_i}{2}-M_B\right)
\left(\left(\frac{E_f+E_i-\omega_f-\omega_i}{2}\right)^2-p_f^2-p_i^2\right) 
-\frac{M_f+M_i}{2}M_fM_i
\right. \nonumber \\&&
-\frac{M_f^2+M_i^2}{2}M_B\   
-\frac{W_f+W_i+M_f+M_i}{2}\left(\left(\frac{E_f+E_i-\omega_f-\omega_i}{2}\right)^2
-p_f^2-p_i^2
+\left(M_f+M_i\right)M_B
\right. \nonumber \\ && \left. \left.
+M_fM_i\vphantom{\left(\left(\frac{E_f}{2}\right)^2\right)}\right)\right]\ ,  \nonumber \\
Y^{(SO)}_B&=&\frac{f_{14}f_{23}}{m_{\pi^+}^2}\ N^-_{fi} \left[-\frac{M_f+M_i}{2}-M_B+\frac{W_f+W_i+M_f+M_i}{2}\right]2p_fp_i \ . \nonumber \\
\end{eqnarray}

\begin{center}
  {\bf (ii) $J^P=\frac{1}{2}^+$ pole term} \\
\end{center}
\label{app:I.6}
(i) \underline{pseudoscalar coupling:}
\begin{eqnarray}
X^{(C)}_B&=&-g_{12}g_{34}\ N^+_{fi} \left[M_B-\frac{W_f+W_i}{2}\right]\ , \nonumber \\
Y^{(C)}_B&=&g_{12}g_{34}\ N^-_{fi} \left[M_B+\frac{W_f+W_i}{2}\right]\ , \nonumber \\
X^{(SO)}_B&=&-g_{12}g_{34}\ N^-_{fi} \left[M_B+\frac{W_f+W_i}{2}\right]\ .
\end{eqnarray}

\noindent (ii) \underline{pseudovector coupling:}
\begin{eqnarray}
X^{(C)}_B&=&\frac{f_{12}f_{34}}{m^2_{\pi^+}}\ N^+_{fi}
\left[\vphantom{\frac{M_f^2}{2}} \left(-\frac{M_f+M_i}{2}-M_B\right)s
+\frac{M_f+M_i}{2}M_fM_i
+\frac{W_f+W_i-M_f-M_i}{2}\left[\vphantom{\left(M_i\right)M_BM_f}s+\left(M_f+
\right. \right. \right. \nonumber \\ && \left. \left. \left.
M_i\right)M_B
+M_fM_i\right]
+\frac{M_f^2+M_i^2}{2}M_B \right]\ , 
\nonumber \\
Y^{(C)}_B&=&\frac{f_{12}f_{34}}{m^2_{\pi^+}}\ N^-_{fi}
\left[\vphantom{\frac{M_f^2}{2}} -\left(-\frac{M_f+M_i}{2}-M_B\right)s
-\frac{M_f+M_i}{2}M_fM_i
+\frac{W_f+W_i-M_f-M_i}{2}\left[\vphantom{\left(M_i\right)M_BM_f}s+\left(M_f+
\right. \right. \right. \nonumber \\ && \left. \left. \left.
M_i\right)M_B
+M_fM_i\right] 
-\frac{M_f^2+M_i^2}{2}M_B \right]\ , \nonumber \\
X^{(SO)}_B&=&-\frac{f_{12}f_{34}}{m^2_{\pi^+}}\ N^-_{fi}
\left[\vphantom{\frac{M_f^2}{2}} -\left(-\frac{M_f+M_i}{2}-M_B\right)s
-\frac{M_f+M_i}{2}M_fM_i
+\frac{W_f+W_i-M_f-M_i}{2}\left[\vphantom{\left(M_i\right)M_BM_f}s+\left(M_f+
\right. \right. \right. \nonumber \\ && \left. \left. \left.
M_i\right)M_B
+M_fM_i\right] 
-\frac{M_f^2+M_i^2}{2}M_B \right]\ .
\end{eqnarray}

\begin{center}
  {\bf (iii) $J^P=\frac{1}{2}^-$ baryon-exchange} \\
\end{center}
\label{app:I.3a}
(i) \underline{scalar coupling:}
\begin{eqnarray}
X^{(C)}_B&=&g^{*(s)}_{14}g^{*(s)}_{23} N^+_{fi} \left[-M_B+\frac{W_f+W_i}{2}-M_f-M_i\right]\ , \nonumber \\
Y^{(C)}_B&=&g^{*(s)}_{14}g^{*(s)}_{23}\ N^-_{fi} \left[M_B+\frac{W_f+W_i}{2}+M_f+M_i\right]\ , \nonumber \\
X^{(SO)}_B&=&-g^{*(s)}_{14}g^{*(s)}_{23}\ N^-_{fi} \left[M_B+\frac{W_f+W_i}{2}+M_f+M_i\right] .
\end{eqnarray}

\noindent (ii) \underline{vector coupling:}
\begin{eqnarray}
X^{(C)}_B&=&-\frac{f^{*(v)}_{14}f^{*(v)}_{23}}{m_{\pi^+}^2}\ N^+_{fi}
\left[\left(-\frac{M_f+M_i}{2}+M_B\right)
\left(\left(\frac{E_f+E_i-\omega_f-\omega_i}{2}\right)^2-p_f^2-p_i^2\right)
+\frac{M_f+M_i}{2}M_fM_i
\right. \nonumber \\ &&
-\frac{M_f^2+M_i^2}{2}M_B
-\frac{W_f+W_i-M_f-M_i}{2}\left( \left(\frac{E_f+E_i-\omega_f-\omega_i}{2}\right)^2
-p_f^2-p_i^2 
-\left(M_f+M_i\right)M_B
\right. \nonumber \\ && \left. \left.
+M_fM_i\vphantom{\left(\left(\frac{E_f}{2}\right)^2\right)} \right)\right]\ , \nonumber \\
Y^{(C)}_B&=&-\frac{f^{*(v)}_{14}f^{*(v)}_{23}}{m_{\pi^+}^2}\ \left[\vphantom{\left(\left(\frac{E_f}{2}\right)^2\right)}N^+_{fi}
\left[-\left(-\frac{M_f+M_i}{2}+M_B\right)
+\frac{W_f+W_i-M_f-M_i}{2}\right]2p_fp_i 
+N^-_{fi}
\left[\vphantom{\left(\left(\frac{E_f}{2}\right)^2\right)}-\left(-\frac{M_f+M_i}{2}
\right. \right. \right. \nonumber\\ && \left.
+M_B\right)\left(\left(\frac{E_f+E_i-\omega_f-\omega_i}{2}\right)^2-p_f^2-p_i^2\right)
-\frac{M_f+M_i}{2}M_fM_i+\frac{M_f^2+M_i^2}{2}M_B
\nonumber  \\&&\left.\left.
-\frac{W_f+W_i+M_f+M_i}{2}
\left(\left(\frac{E_f+E_i-\omega_f-\omega_i}{2}\right)^2-p_f^2-p_i^2
-\left(M_f+M_i\right)M_B+M_fM_i\right)\right]\right]\ ,  
\nonumber \\
Z^{(C)}_B&=&-\frac{f^{*(v)}_{14}f^{*(v)}_{23}}{m_{\pi^+}^2}\ N^-_{fi} \left[-\frac{M_f+M_i}{2}+M_B+\frac{W_f+W_i+M_f+M_i}{2}\right]2p_fp_i \ , \nonumber \\
X^{(SO)}_B&=&\frac{f^{*(v)}_{14}f^{*(v)}_{23}}{m_{\pi^+}^2}\ N^-_{fi}
\left[-\left(-\frac{M_f+M_i}{2}+M_B\right)
\left(\left(\frac{E_f+E_i-\omega_f-\omega_i}{2}\right)^2-p_f^2-p_i^2\right) 
-\frac{M_f+M_i}{2}M_fM_i
\right. \nonumber \\&&
+\frac{M_f^2+M_i^2}{2}M_B\   
-\frac{W_f+W_i+M_f+M_i}{2}\left(\left(\frac{E_f+E_i-\omega_f-\omega_i}{2}\right)^2
-p_f^2-p_i^2
-\left(M_f+M_i\right)M_B
\right. \nonumber\\ && \left. \left.
+M_fM_i\vphantom{\left(\left(\frac{E_f}{2}\right)^2\right)}\right)\right]\ ,  \nonumber \\
Y^{(SO)}_B&=&\frac{f^{*(v)}_{14}f^{*(v)}_{23}}{m_{\pi^+}^2}\ N^-_{fi} \left[-\frac{M_f+M_i}{2}+M_B+\frac{W_f+W_i+M_f+M_i}{2}\right]2p_fp_i \ . 
\end{eqnarray}

\begin{center}
  {\bf (iv) $J^P=\frac{1}{2}^-$ pole term} \\
\end{center}
\label{app:I.6a}
(i) \underline{scalar coupling:}
\begin{eqnarray}
X^{(C)}_B&=&g^{*(v)}_{12}g^{*(v)}_{34}\ N^+_{fi} \left[M_B+\frac{W_f+W_i}{2}\right]\ , \nonumber \\
Y^{(C)}_B&=&g^{*(v)}_{12}g^{*(v)}_{34}\ N^-_{fi} \left[-M_B+\frac{W_f+W_i}{2}\right]\ , \nonumber \\
X^{(SO)}_B&=&-g^{*(v)}_{12}g^{*(v)}_{34}\ N^-_{fi} \left[-M_B+\frac{W_f+W_i}{2}\right]\ .
\end{eqnarray}

\noindent (ii) \underline{vector coupling:}
\begin{eqnarray}
X^{(C)}_B&=&\frac{f^{*(v)}_{12}f^{*(v)}_{34}}{m^2_{\pi^+}}\ N^+_{fi}
\left[\vphantom{\frac{M_f^2}{2}} \left(-\frac{M_f+M_i}{2}+M_B\right)s
+\frac{M_f+M_i}{2}M_fM_i
+\frac{W_f+W_i-M_f-M_i}{2}\left[s-\left(M_f+
\right. \right. \right. \nonumber \\ && \left. \left. \left.
M_i\right)M_B
+M_fM_i\vphantom{\left(M_f\right)}\right] 
-\frac{M_f^2+M_i^2}{2}M_B \right]\ , \nonumber \\
Y^{(C)}_B&=&\frac{f^{*(v)}_{12}f^{*(v)}_{34}}{m^2_{\pi^+}}\ N^-_{fi}
\left[\vphantom{\frac{M_f^2}{2}} -\left(-\frac{M_f+M_i}{2}+M_B\right)s
-\frac{M_f+M_i}{2}M_fM_i
+\frac{W_f+W_i-M_f-M_i}{2}\left[s-\left(M_f+
\right. \right. \right. \nonumber\\ && \left. \left. \left.
M_i\right)M_B
+M_fM_i\vphantom{\left(M_f\right)}\right]
+\frac{M_f^2+M_i^2}{2}M_B \right]\ , \nonumber \\
X^{(SO)}_B&=&-\frac{f^{*(v)}_{12}f^{*(v)}_{34}}{m^2_{\pi^+}}\ N^-_{fi}
\left[\vphantom{\frac{M_f^2}{2}} -\left(-\frac{M_f+M_i}{2}+M_B\right)s
-\frac{M_f+M_i}{2}M_fM_i
+\frac{W_f+W_i-M_f-M_i}{2}\left[s-\left(M_f+
\right. \right. \right. \nonumber \\ && \left. \left. \left.
M_i\right)M_B
+M_fM_i\vphantom{\left(M_f\right)}\right]
+\frac{M_f^2+M_i^2}{2}M_B \right]\ .
\end{eqnarray}

\begin{center}
  {\bf (v) $J^P=\frac{3}{2}^+$ baryon-exchange} \\
\end{center}
\label{app:I.9}
\begin{eqnarray}
X^{(C)}_{Y^*}&=&-\frac{f^*_{14}f^*_{23}}{m_{\pi ^+}^2} N^+_{fi} \left[A_0+\frac{B_0}{2}\left(W_f+W_i-M_i-M_f\right)\right]\ , \nonumber \\
Y^{(C)}_{Y^*}&=&-\frac{f^*_{14}f^*_{23}}{m_{\pi ^+}^2}\left[ N^+_{fi} \left[A_1+\frac{B_1}{2}\left(W_f+W_i-M_f-M_i\right)\right]
+ N^-_{fi} \left[-A_0+\frac{B_0}{2}\left(W_f+W_i+M_f+M_i\right)\right]\right]\ , \nonumber \\
Z^{(C)}_{Y^*}&=&-\frac{f^*_{14}f^*_{23}}{m_{\pi ^+}^2}\left[ N^+_{fi} \left[A_2+\frac{B_2}{2}\left(W_f+W_i-M_f-M_i\right)\right]
+ N^-_{fi} \left[-A_1+\frac{B_1}{2}\left(W_f+W_i+M_f+M_i\right)\right]\right]\ , \nonumber \\
U^{(C)}_{Y^*}&=&\frac{f^*_{14}f^*_{23}}{m_{\pi ^+}^2} N^-_{fi} \left[-A_2+\frac{B_2}{2}\left(W_f+W_i+M_f+M_i\right)\right]\ , \nonumber \\
X^{(SO)}_{Y^*}&=&\frac{f^*_{14}f^*_{23}}{m_{\pi ^+}^2} N^-_{fi} \left[-A_0+\frac{B_0}{2}\left(W_f+W_i+M_i+M_f\right)\right]\ , \nonumber \\
Y^{(SO)}_{Y^*}&=&\frac{f^*_{14}f^*_{23}}{m_{\pi ^+}^2} N^-_{fi} \left[-A_1+\frac{B_1}{2}\left(W_f+W_i+M_i+M_f\right)\right]\ , \nonumber \\
Z^{(SO)}_{Y^*}&=&\frac{f^*_{14}f^*_{23}}{m_{\pi ^+}^2} N^-_{fi} \left[-A_2+\frac{B_0}{2}\left(W_f+W_i-M_f-M_i\right)\right] .
\end{eqnarray}
Where $A_0$, $A_1$, $A_2$, $B_0$, $B_1$ and $B_2$ depend on the mass and momentum of the particles as follows.
\begin{eqnarray}
A_0&=&\frac{1}{12M_{Y^*}^2}\left(M_f+M_i\right)\left(-2p_fp_iz_u+M_{Y^*}^2\right)^2 
+\left(-\frac{1}{6M_{Y^*}^2}\left(M_f^2+M_i^2-m_f^2-m_i^2\right)\frac{M_f+M_i}{2}+\frac{M_{Y^*}}{3}+
\right. \nonumber \\ &&
\frac{M_f+M_i}{6} 
\left.
-\frac{1}{6M_{Y^*}}\left(\frac{M_i}{2}\left(M_i-M_f\right)-m_f^2+\frac{M_f}{2}\left(M_f-M_i\right)-m_i^2\right)\right)\left(-2p_fp_iz_u+M_{Y^*}^2\right) \nonumber \\
&&+\frac{1}{2}\left(\frac{M_f+M_i}{2}+M_{Y^*}\right)\left(\frac{\left(E_f-E_i\right)^2+\left(\omega_f-\omega_i\right)^2}{2}-p_f^2-p_i^2\right) 
-\frac{m_f^2+m_i^2}{2}\left(\frac{M_f+M_i}{2}+M_{Y^*}\right)
\nonumber \\&&
+\frac{1}{12M_{Y^*}^2}\left(M_f+M_i\right)\left(M_f^2-m_i^2\right)
\left(M_i^2-m_f^2\right) 
-\frac{M_{Y^*}}{6}\left(M_f^2+M_i^2\right)+\frac{1}{12}\left(\vphantom{\left(M_f^2\right)}\left(m_i^2-m_f^2\right)\left(M_f-M_i\right)
\right. 
\nonumber \\ &&\left.
-\left(M_f^2+M_i^2\right)
\left(M_f+M_i\right)\right)
+\frac{1}{6M_{Y^*}}\left(\left(M_f^2-m_i^2\right)\left(\frac{M_i}{2}\left(M_i-M_f\right)-m_f^2\right)+\left(M_i^2-m_f^2\right)
\times\right.\nonumber \\&&\left.
\left(\frac{M_f}{2}\left(M_f-M_i\right)-m_i^2\right)\right)\ ,  \nonumber
\end{eqnarray}
\begin{eqnarray}
A_1&=&\left[-\frac{1}{12M_{Y^*}^2}\left(M_f+M_i\right)2\left(-2p_fp_iz_u+M_{Y^*}^2\right) 
+\frac{1}{6M_{Y^*}^2}\left(M_f^2+M_i^2-m_f^2-m_i^2\right)\frac{M_f+M_i}{2}-\frac{M_{Y^*}}{3}-
\right. \nonumber \\ &&\left.
\frac{M_f+M_i}{6} 
+\frac{1}{6M_{Y^*}}\left(\frac{M_i}{2}\left(M_i-M_f\right)-m_f^2+\frac{M_f}{2}\left(M_f-M_i\right)-m_i^2\right) 
+\frac{1}{2}\left(\frac{M_f+M_i}{2}+M_{Y^*}\right)\right] 2p_fp_i\ , \nonumber \\
A_2&=&\frac{1}{12M_{Y^*}^2}\left(M_f+M_i\right)\left(2p_fp_i\right)^2 \ ,\nonumber \\
B_0&=&\frac{m_f^2+m_i^2}{2}-\frac{1}{6M_{Y^*}^2}\left(M_f^2-m_i^2\right)\left(M_i^2-m_f^2\right)+\frac{M_{Y^*}}{3}\left(M_f+M_i\right) 
-\frac{m_f^2+m_i^2-\left(M_f+M_i\right)^2}{6}
\nonumber \\&&
+\frac{1}{6M_{Y^*}}\left(M_f\left(M_i^2-m_f^2\right)+M_i\left(M_f^2-m_i^2\right)\right) 
-\frac{1}{2}\left(\frac{\left(E_f-E_i\right)^2+\left(\omega_f-\omega_i\right)^2}{2}-p_f^2-p_i^2\right)
\nonumber \\&&
-\frac{1}{6M_{Y^*}^2}\left(-2p_fp_iz_u+M_{Y^*}^2\right)^2 
+\left(\frac{1}{6M_{Y^*}^2}\left(M_f^2+M_i^2-m_f^2-m_i^2\right)-\frac{1}{6M_{Y^*}}\left(M_f+M_i\right)\right)
\times \nonumber \\&&
\left(-2p_fp_iz_u+M_{Y^*}^2\right)\ ,  \nonumber \\
B_1&=&\left[\frac{1}{3M_{Y^*}^2}\left(-2p_fp_iz_u+M_{Y^*}^2\right)
-\frac{1}{6M_{Y^*}}\left(\frac{1}{M_{Y^*}}\left(M_f^2+M_i^2-m_f^2-m_i^2\right)-\left(M_f+M_i\right)\right)-\frac{1}{2}\right]2p_fp_i\ ,  \nonumber \\
B_2&=&-\frac{1}{6M_{Y^*}^2}\left(2p_fp_i\right)^2\ .
\end{eqnarray}

\begin{center}
  {\bf (vi) $J^P=\frac{3}{2}^+$ pole term} \\
\end{center}
\label{app:I.10}
\begin{eqnarray}
X^{(C)}_{Y^*}&=&\frac{f^*_{12}f^*_{34}}{m_{\pi ^+}^2} N^+_{fi} \left[A_0+\frac{B_0}{2}\left(W_f+W_i-M_i-M_f\right)\right]\ , \nonumber \\
Y^{(C)}_{Y^*}&=&\frac{f^*_{12}f^*_{34}}{m_{\pi ^+}^2}\left[ N^+_{fi} \left[A_1+\frac{B_1}{2}\left(W_f+W_i-M_f-M_i\right)\right]
+ N^-_{fi} \left[-A_0+\frac{B_0}{2}\left(W_f+W_i+M_f+M_i\right)\right]\right]\ , \nonumber \\
Z^{(C)}_{Y^*}&=&\frac{f^*_{12}f^*_{34}}{m_{\pi ^+}^2} N^-_{fi} \left[-A_1+\frac{B_1}{2}\left(W_f+W_i+M_f+M_i\right)\right]\ , \nonumber \\
X^{(SO)}_{Y^*}&=&-\frac{f^*_{12}f^*_{34}}{m_{\pi ^+}^2} N^-_{fi} \left[-A_0+\frac{B_0}{2}\left(W_f+W_i+M_i+M_f\right)\right]\ , \nonumber \\
Y^{(SO)}_{Y^*}&=&-\frac{f^*_{12}f^*_{34}}{m_{\pi ^+}^2} N^-_{fi} \left[-A_1+\frac{B_1}{2}\left(W_f+W_i+M_i+M_f\right)\right]\ .\nonumber \\
&&
\end{eqnarray}
Where $A_0$, $A_1$, $B_0$ and $B_1$ depend on the mass and momentum of the particles as follows.
\begin{eqnarray}
A_0&=&\frac{1}{2}\left(\frac{\left(E_f-E_i\right)^2+\left(\omega_f-\omega_i\right)^2}{2}-p_f^2-p_i^2-m_f^2-m_i^2\right)\left(\frac{M_f+M_i}{2}+M_{Y^*}\right)
+\frac{1}{6M_{Y^*}^2}\left(M_f^2-m_f^2-s\right)
\times
\nonumber 
\\&&
\left(M_i^2-m_i^2-s\right)\frac{M_f+M_i}{2} 
+ \frac{1}{3 M_{Y^*}}\left(s-\frac{1}{2}\left(M_f^2
+M_i^2\right)\right)
+\frac{1}{3}\left(\frac{M_f+M_i}{2} s -\frac{\left(M_f^2+M_i^2\right)\left(M_f+M_i\right)}{4}
\right. \nonumber\\ &&\left.
+\frac{m_f^2-m_i^2}{4}
\left(M_f-M_i\right)\right) 
+\frac{1}{6M_{Y^*}}
\left(\left(M_f^2-m_f^2-s\right)\left(\frac{M_i}{2}\left(M_i-M_f\right)-m_i^2\right)-
\left(M_i^2-m_i^2-s\right)
\times \right.\nonumber \\ &&\left.
\left(\frac{M_f}{2}\left(M_i-M_f\right)+m_f^2\right)\right)\ ,  \nonumber \\
A_1&=&\frac{1}{2}\left[\frac{M_f+M_i}{2}+M_{Y^*}\right]2p_fp_i\ , \nonumber \\
B_0&=&\frac{1}{2}\left(\frac{\left(E_f-E_i\right)^2+\left(\omega_f-\omega_i\right)^2}{2}-p_f^2-p_i^2-m_f^2-m_i^2 \right)
-\frac{1}{6M_{Y^*}^2}\left(s-M_f^2+m_f^2\right)
\left(s-M_i^2+m_i^2\right)
\nonumber \\&&
-\frac{M_{Y^*}}{3}\left(M_f+M_i\right)+\frac{m_f^2+m_i^2-\left(M_f+M_i\right)^2}{6} 
-\frac{1}{6M_{Y^*}}
\left(M_f\left(M_i^2-m_i^2-s\right)+M_i\left(M_f^2-m_f^2-s\right)\right)\ ,  \nonumber \\
B_1&=&p_fp_i\ .
\end{eqnarray}

\subsection{Meson-exchange}
\label{app:I.1}
\begin{center}
  {\bf (i) $J^P=0^{++}$ scalar-meson-exchange} \\
\end{center}
\begin{eqnarray}
X^{(C)}_S&=&-g_{PPS}g_{S}\ N^+_{fi}\ , \nonumber \\
Y^{(C)}_S&=&g_{PPS}g_{S}\ N^-_{fi}\ , \nonumber \\
X^{(SO)}_S&=&-g_{PPS}g_{S}\ N^-_{fi}\ .
\end{eqnarray}

\begin{center}
  {\bf (ii) $J^P=1^{--}$ vector-meson-exchange} \\
\end{center}
\begin{eqnarray}
X^{(C)}_V&=&-g_{PPV}g_V\ N^+_{fi}
\left[\frac{\left(m_f^2-m_i^2\right)\left(M_i-M_f\right)}{m_V^2}-\left(W_f+W_i
-M_i-M_f\right)\right] 
-g_{PPV}f_V\ N^+_{fi}
\left[\vphantom{\frac{\left(E_f\right)p_f^2}{2{\cal M}}}-\frac{M_f+M_i}{2{\cal M}}
\right. \nonumber \\ &&\times \left.
\left(W_f+W_i
-M_i-M_f\right)+\frac{\left(\omega _f+\omega _i\right)\left(E_f+E_i\right)+p_f^2+p_i^2}{2{\cal M}}\right]\ ,  \nonumber \\
Y^{(C)}_V&=&-g_{PPV}g_V\ N^-_{fi}
\left[-\frac{\left(m_f^2-m_i^2\right)\left(M_i-M_f\right)}{m_V^2}-\left(W_f+W_i
+M_i+M_f\right)\right] 
-g_{PPV}f_V\ \left[\vphantom{\frac{\left(E_f\right)p_f^2}{2{\cal M}}} N^+_{fi} \ \frac{p_fp_i}{{\cal M}}
\right. \nonumber \\&&\left.
+N^-_{fi} \left[-\frac{M_f+M_i}{2{\cal M}}\left(W_f+W_i+M_i+M_f\right)
-\frac{\left(\omega _f+\omega _i\right)\left(E_f+E_i\right)+p_f^2+p_i^2}{2{\cal M}}\right]\right]\ ,  \nonumber \\
Z^{(C)}_V&=&g_{PPV}f_V\ N^-_{fi} \ \frac{p_fp_i}{{\cal M}}\ , \nonumber \\
X^{(SO)}_V&=&g_{PPV}g_V\ N^-_{fi}
\left[-\frac{\left(m_f^2-m_i^2\right)\left(M_i-M_f\right)}{m_V^2}-\left(W_f+W_i
+M_f+M_i\right)\right] 
+g_{PPV}f_V\ N^-_{fi}
\left[\vphantom{\frac{\left(E_f\right)p_f^2}{2{\cal M}}}-\frac{M_f+M_i}{2{\cal M}}
\right. \nonumber \\ && \left. \times
\left(W_f+W_i
+M_f+M_i\right)-\frac{\left(\omega _f+\omega _i\right)\left(E_f+E_i\right)+p_f^2+p_i^2}{2{\cal M}}\right]\ ,  \nonumber \\
Y^{(SO)}_V&=&-g_{PPV}f_V\ N^-_{fi} \ \frac{p_fp_i}{{\cal M}}\ .
\end{eqnarray}

\begin{center}
  {\bf (iii) $J^P=2^{++}$ tensor-meson-exchange} \\
\end{center}
\begin{eqnarray}
X^{(C)}_T&=&-\frac{g_{PPT}F_1}{m_{\pi^+}} N^+_{fi} \left[A_0+\frac{B_0}{2}\left(W_f+W_i-M_i-M_f\right)\right]\ , \nonumber \\
Y^{(C)}_T&=&-\frac{g_{PPT}F_1}{m_{\pi^+}}\left[N^+_{fi} \left[A_1+\frac{B_1}{2}\left(W_f+W_i-M_f-M_i\right)\right]
+N^-_{fi} \left[-A_0+\frac{B_0}{2}\left(W_f+W_i+M_f+M_i\right)\right]\right]\ , \nonumber \\
Z^{(C)}_T&=&-\frac{g_{PPT}F_1}{m_{\pi^+}}\left[N^+_{fi} A_2
+N^-_{fi} \left[-A_1+\frac{B_1}{2}\left(W_f+W_i+M_f+M_i\right)\right]\right]\ , \nonumber \\
U^{(C)}_T&=&\frac{g_{PPT}F_1}{m_{\pi^+}} N^-_{fi} A_2\ , \nonumber \\
X^{(SO)}_T&=&\frac{g_{PPT}F_1}{m_{\pi^+}} N^-_{fi} \left[-A_0+\frac{B_0}{2}\left(W_f+W_i+M_i+M_f\right)\right]\ , \nonumber \\
Y^{(SO)}_T&=&\frac{g_{PPT}F_1}{m_{\pi^+}} N^-_{fi} \left[-A_1+\frac{B_1}{2}\left(W_f+W_i+M_i+M_f\right)\right]\ , \nonumber \\
Z^{(SO)}_T&=&-\frac{g_{PPT}F_1}{m_{\pi^+}} N^-_{fi} A_2\ .
\end{eqnarray}
Where $A_0$, $A_1$, $A_2$, $B_0$ and $B_1$ depend on the mass and momentum of the particles as follows.
\begin{eqnarray}
A_0&=&\frac{F_2}{F_1}\left[\vphantom{\frac{\left( M_f^2\right)^2}{4m_T^2}} \left(\frac{\left(\omega _i+\omega_f\right)\left(E_f+E_i\right)+p_f^2+p_i^2}{2}\right)^2 +\frac{1}{4m_T^4}\left(m_f^2-m_i^2\right)^2\left(M_i^2-M_f^2\right)^2 
\right. \nonumber \\&&
-\frac{\left(\omega _i+\omega_f\right)\left(E_f+E_i\right)+p_f^2+p_i^2 }{2m_T^2}
\left(m_f^2-m_i^2\right)\left(M_i^2-M_f^2\right) 
-\frac{1}{3}
\left(-2\left(m_f^2+m_i^2\right)+\frac{\left(m_f^2-m_i^2\right)^2}{m_T^2}
\right. \nonumber \\ && \left.
+\frac{\left(E_f-E_i\right)^2+\left(\omega _f-\omega _i\right)^2}{2}-p_f^2-p_i^2\vphantom{\frac{\left( m_f^2\right)^2}{m_T^2}}\right)
\left(\frac{\left(M_i^2-M_f^2\right)^2}{4m_T^2} -\frac{M_f^2+M_i^2}{2}
+\frac{1}{4}\left(\vphantom{\frac{\left(E_f\right)^2}{2}} -p_f^2-p_i^2+
\right. \right. \nonumber \\ && \left. \left. \left.
\frac{\left(E_f-E_i\right)^2+\left(\omega _f-\omega _i\right)^2}{2} \right)\vphantom{\frac{\left( M_f^2\right)^2}{4m_T^2}}\right)\right]
\nonumber \\&&
+\left[\vphantom{\frac{\left( m_f^2\right)^2}{m_T^2}} -\frac{\left(\omega _i+\omega_f\right)\left(E_f+E_i\right)+p_f^2+p_i^2}{2m_T^2}\left(m_f^2-m_i^2\right)\left(M_i-M_f\right) 
\right.\nonumber \\ &&
+\frac{1}{2m_T^4}\left(m_f^2-m_i^2\right)^2\left(M_i^2-M_f^2\right)\left(M_i-M_f\right) 
-\frac{1}{3}
\left(-2\left(m_f^2+m_i^2\right)+\frac{\left(m_f^2-m_i^2\right)^2}{m_T^2}-p_f^2-p_i^2
\right. \nonumber \\ && \left.\left.
+\frac{\left(E_f-E_i\right)^2+\left(\omega _f-\omega _i\right)^2}{2}\vphantom{\frac{\left( m_f^2\right)^2}{m_T^2}}\right)
\left(\frac{\left(M_i^2-M_f^2\right)}{2m_T^2}\left(M_i-M_f\right) -\frac{1}{2}\left(M_f+M_i\right)\right)\right]\ ,\nonumber \\
A_1&=&\frac{F_2}{F_1}\left[\left(\omega _i+\omega_f\right)\left(E_f+E_i\right)+p_f^2+p_i^2-\frac{\left(m_f^2-m_i^2\right)\left(M_i^2-M_f^2\right)}{m_T^2} 
-\frac{1}{6}\left(-2\left(m_f^2+m_i^2\right)+\frac{\left(m_f^2-m_i^2\right)^2}{m_T^2}
\right. \right. \nonumber \\ && \left.
+\frac{\left(E_f-E_i\right)^2+\left(\omega _f-\omega _i\right)^2}{2}-p_f^2-p_i^2\vphantom{\frac{\left( m_f^2\right)^2}{m_T^2}}\right) 
-\frac{\left(M_i^2-M_f^2\right)^2}{6m_T^2}+\frac{M_f^2+M_i^2}{3} 
-\frac{1}{6}\left(\frac{\left(E_f-E_i\right)^2+\left(\omega _f-\omega _i\right)^2}{2}
\right. \nonumber\\ && \left.\left.
-p_f^2-p_i^2\vphantom{\frac{\left( E_f\right)^2}{2}}\right)\vphantom{\frac{\left( m_f^2\right)^2}{m_T^2}}\right] p_fp_i 
\nonumber \\&&
+\left[-\frac{\left(m_f^2-m_i^2\right)\left(M_i-M_f\right)}{m_T^2}-\frac{\left(M_i^2-M_f^2\right)\left(M_i-M_f\right)}{3m_T^2}+
\frac{M_f+M_i}{3}\right]p_fp_i \ ,\nonumber \\
A_2&=&-\frac{1}{3}p_f^2p_i^2\frac{F_2}{F_1}\ , \nonumber \\
B_0&=&\left(\omega _i+\omega_f\right)\left(E_f+E_i\right)+p_f^2+p_i^2-\frac{\left(m_f^2-m_i^2\right)\left(M_i^2-M_f^2\right)}{m_T^2}\ , \nonumber \\
B_1&=&2p_fp_i\ .
\end{eqnarray}

\subsection{Pomeron-exchange}
\label{app:I.8}
\begin{eqnarray}
X^{(C)}_P&=&g_{PPP}g_{P}\ N^+_{fi} \ , \nonumber \\
Y^{(C)}_P&=&-g_{PPP}g_{P}\ N^-_{fi} \ , \nonumber \\
X^{(SO)}_P&=&g_{PPP}g_{P}\ N^-_{fi} \ .
\end{eqnarray}

\end{widetext}

\end{document}